\documentclass[referee]{aa} 
\usepackage{graphicx}
\usepackage{txfonts}
\def\cm{\ifmmode {\rm cm}^{-1} \else cm$^{-1}$ \fi}
\def\s{\ifmmode {\rm s}^{-1} \else s$^{-1}$ \fi}
\def\cc{\ifmmode {\rm cm}^{-3} \else cm$^{-3}$ \fi}
\def\cs{\ifmmode {\rm cm}^{-2} \else cm$^{-2}$ \fi}
\def\g{\ifmmode \gamma \else $\gamma$\fi}
\def\G{\ifmmode \Gamma \else $\Gamma$\fi}
\def\Gs{\ifmmode \Gamma~ \else $\Gamma~$\fi}

\def\gc{\ifmmode \gamma_{\rm c} \else $\gamma_{\rm c}$ \fi}

\def\gsim{\mathrel{\raise.5ex\hbox{$>$}\mkern-14mu
             \lower0.6ex\hbox{$\sim$}}}
\def\lsim{\mathrel{\raise.3ex\hbox{$<$}\mkern-14mu
             \lower0.6ex\hbox{$\sim$}}}
\def\simless{\mathbin{\lower 3pt\hbox
     {$\rlap{\raise 5pt\hbox{$\char'074$}}\mathchar"7218$}}}   
\def\simmore{\mathbin{\lower 3pt\hbox
     {$\rlap{\raise 5pt\hbox{$\char'076$}}\mathchar"7218$}}}   
\def\4u{4U 1728--34}

\begin{document}
   \title{QPOs in the Time Domain: An Autocorrelation Analysis}

\titlerunning{QPOs in Autocorrelation Functions}
\authorrunning{Fukumura et al.}


   \author{K. Fukumura
          \inst{1,2}, C. R. Shrader\inst{1,3}, J. W. Dong\inst{4},
          \and
          D. Kazanas\inst{1}
          }

   \institute{Astrophysics Science Division, NASA/Goddard Space
              Flight Center, Greenbelt, MD 20771\\
              \email{Demos.Kazanas@nasa.gov}
              \and
              University of Maryland, Baltimore County
             (UMBC/CRESST), Baltimore, MD 21250\\
             \email{Keigo.Fukumura@nasa.gov}
             \and
              University Space Research Association, 10211
             Wincopin Circle, Suite 620, Columbia, MD 21044\\
             \email{Chris.R.Shrader@nasa.gov}
             \and
              Columbia University, Department of Mechanical Engineering, New York, NY 10027\\
             \email{jwd9188@gmail.com}}



  \abstract
{Motivated by the recent proposal that one can obtain quasi-periodic
oscillations (QPOs) by photon echoes
%
%
manifesting as non-trivial features
in the autocorrelation function (ACF), we study the ACFs of the light curves of three accreting black hole candidates and a
neutron star already known to exhibit QPOs namely,
GRS~1915+105, XTE~J1550-564, XTE~J1859+226 and Cygnus X-2. }
{We present a comparative study of the timing properties of these
systems in the frequency and time domain in search for
similarities/differences that may provide clues to the physics
underlying the QPO phenomenon. 
 }
{We compute and focus on the form of the ACFs in search of
systematics or specific temporal properties at the time scales associated
with the known QPO frequencies in comparison with the corresponding PDS. }
{Even within our small object sample we find both similarities as
well as significant and subtle differences  in the form of the ACFs
both amongst black holes and between black holes and neutron stars
to warrant a closer look at the QPO phenomenon in the time domain:
The QPO features manifest as an oscillatory behavior of the ACF at
lags near zero; the oscillation damps exponentially on time scales
equal to the inverse QPO width to a level of a percent or so. In
black holes this oscillatory behavior is preserved and easily
discerned at much longer lags while this is not the case for the
neutron star system Cyg X-2. The ACF of GRS 1915+105 provides an
exception to this general behavior in that its decay is {\em linear
\em} in time indicating an undamped oscillation of coherent phase.
We present simple {\em ad hoc \em} models that reproduce these
diverse time domain behaviors and we speculate that their origin is
the phase coherence of the underlying oscillation.}
{It appears plausible that time domain analyses, {\em complementary
\em} to the more common frequency domain ones, could impose tighter
constraints and provide clues for the driving mechanisms behind the
QPO phenomenon. }

\keywords{black hole physics --- binaries: general ---
stars: individual (Cygnus X-2, GRS~1915+105, XTE~J1550-564, XTE~J1859+226) --- stars: oscillations --- X-rays: stars}

   \maketitle
%

\section{Introduction}

In the past several decades the study of accretion powered sources,
i.e. neutron stars and black holes, whether at the galactic or
extragalactic scales, has been one of the central themes of high
energy astrophysics. However, the small angular size of these
systems puts them beyond the imaging capabilities of present and perhaps
foreseeable future telescopes. Therefore, by necessity, the study of
the structure and dynamics of the accretion flows responsible for
their observed luminosity is effected through studies of their
spectra and variability.

For the galactic sources in particular, most of their luminosity is
emitted at X-ray energies ($E \sim 0.1 - 20$ keV) and their spectra
at their highest energies, $E \gsim 5$ keV, are thought to be
produced by the Comptonization process, while at lower energies they
exhibit a quasi-thermal component, usually fit by the spectrum of a
multi-color disk (MCD).
However, spectral studies alone, in particular of spectra due to
Comptonization, constrain mainly the column density of the hot gas
along the observer's line of sight. This is not sufficient to
provide information about the accreting flow dynamics, which require
in addition knowledge of source size (e.g. Hua, Kazanas \& Cui~1999); this is usually
estimated from the  source variability, thus arguing for the
combined spectro-temporal analysis of the physics of accretion
flows.

The variability of these sources, as conventionally determined
through their power density spectra (PDS), spans many decades in
Fourier frequency at roughly constant power ($\nu \sim 10^{-2} -
10^3$ Hz extending in some cases to frequencies as low as
$10^{-6}$ Hz), presumably an indication that the associated flows
span also a large number of decades in radius. However, in addition
to these broad power spectra of aperiodic variability, the PDS of
accretion powered sources exhibit often excess variability also at
particular narrow frequency bands known as quasi-periodic
oscillations (QPOs). These are presumably associated with
frequencies characteristic of the flow dynamics and/or the radiative
transfer of the photons in the accretion flow.

QPOs are found in the PDS of both accreting galactic black hole
candidates (GBHCs; e.g. Remillard \& McClintock~2006) and neutron
stars in low-mass X-ray binaries (LMXBs; e.g. van der Klis~2004).
Perhaps with exception of the neutron star kHz QPO, their
frequencies are generally lower than those characteristic of the
dynamics of accretion onto an object of size a few Schwarzschild
radii. Roughly twenty five years of QPO observations have produced a
rich phenomenology which to date remains by and large unaccounted
for. The generic QPO model, employing qualitative arguments about
the dynamics or kinematics of the accretion flow and parameters
chosen to reproduce the frequencies already observed, has rather
limited predictive power. In the absence of a general, robust
theoretical framework that it can reproduce the observed
phenomenology it is not even clear whether the physics underlying
the QPOs at the different frequency regimes and different sources
involve similar or fundamentally different processes. Most
interestingly, discerning the corresponding oscillations in the
source light curves (LCs) has remained difficult (particularly more
so in active galactic nuclei; e.g. Vaughan \& Uttley~2005), even
though on occasion the PDS exhibit QPOs with large Q-values ($Q
\simeq 10$, where $Q \equiv \nu_0/\Delta \nu \simeq 10$ and $\Delta
\nu$ is the QPO FWHM); an exception are the light curves that yield
the low frequency ($<0.1$ Hz) QPOs of GRS~1915+105 of
Fig.~\ref{fig:ps}c, in which the corresponding oscillations are
possible to discern in the LC.

It is generally believed that the QPO phenomenon involves an
oscillation of sorts; e.g. orbiting blobs of gas (clumpy hot spots) in the context of general relativistic diskoseismology under strong gravity (e.g. Schnittman~2005). However, while sufficient, this is not
necessary, neither the PDS provides any clues on the shape of the
oscillating signal. Specifically, one could also obtain QPO-like
features in the PDS of a time series, among others, if the latter
involves a given signal and its ``echo", i.e. a well defined lag of
itself (see, e.g., Kazanas \& Hua~1999), which in accretion powered
sources will most likely be the result of a specific geometric
arrangement of the system (e.g a warped disk). We have recently
shown the possibility of (harmonically spaced) ``echo" QPOs due to general relativistic frame-dragging in
the (simulated) inherently aperiodic LCs of rapidly rotating black
holes (i.e. black holes with $a/M \gsim 0.94$ where $a$ is specific
spin and $M$ is mass of a black hole) (Fukumura \&
Kazanas~2008,Fukumura, Kazanas, \& Stephenson~2009).
At this point it is not clear whether any of the observed  GBHCs
exhibit these ``ergospheric" (or any other type of echo) QPOs (there
is a search of the existing data bases going on). However, should
the physics underlying certain harmonically spaced QPOs (often seen
in the PDS of accreting sources) be an echo
%
%
rather than an oscillation, the ACF would provide an indispensable
analysis tool in verifying this because of its unique signature,
namely its double peak structure (harmonically spaced QPOs are
normally attributed to an underlying oscillation of
non-sinusoidal shape).

The timing analysis of accretion powered sources, especially those
that exhibit the QPO phenomenon is typically performed in the
frequency domain through the use of FFT and the corresponding Power
Density Spectra (PDS). However, as argued above, analysis in the
time domain through the ACF may also provide useful,
\emph{complementary\/} information as we will exhibit further on. As
well known, the PDS and ACF are related through a Fourier transform
and one might think that one of them might suffice to obtain all
relevant knowledge about the signal. However, each of these
variability metrics emphasizes different aspects of the underlying
signal and the simultaneous transform of the always present noise
might permit discerning different clues of the process underlying
the QPO phenomenon in the two different domains.

We note that to date analyses of QPO properties the time-domain
(i.e. through ACF) are precious few in the literature (see, e.g.
Fig.~4 in Ebisawa et al.~1989 for the ACF associated with the QPO of
LMC~X-1). Motivated by this fact and the above discussion we have
decided to take a closer look at the ACFs of sources which are known
to exhibit harmonically spaced QPOs at well defined frequencies,
thus shifting the emphasis of the analysis from the frequency to the
time domain.

With these considerations in mind we present in \S 2 the PDS
analysis of the time series of four accretion-powered sources known
to exhibit prominent QPOs with harmonically spaced peaks. Of these,
three are GBHCs and one is a well known accreting LMXB neutron star.
Comparative analysis of the ACFs are then presented in \S 3.
%
%
In \S 4 we present explicit examples of time series with similar PDS
but very different ACFs in order to demonstrate the
\emph{complementary roles\/} of PDS and ACF. We finally summarize
and review these findings and draw some general conclusions in \S 5.

%
%


\section{Source Selection and their PDS Analysis}

The sources used in our analysis were selected by the requirement
that they exhibit large amplitude QPOs at well defined frequencies
and that they represent different types of objects, in order to test
whether their overall morphological properties play a role in the
type of QPOs and ACFs we obtain. The sources we have chosen usually
exhibit QPOs in more than one frequency range, typically at a few Hz
and also at tens to a hundred Hz. Since the lower frequency QPOs are
the more prominent ones, we focus our analysis at these. Actually,
it is not obvious that the ACF analysis we present herein can
discern the higher frequency species of these features, but the
answer to this question will have to wait for a more detailed
analysis.

With the above very general criteria we focused on the following
sources:

{\em 1. XTE~J1550-564:} This is a well known X-ray nova, discovered
by the ASM aboard {\it RXTE} in September 1998 (Smith~1998). The fact
that the source remained active for an extended period afforded the
opportunity to study its spectral and temporal properties as a
function of the source flux. It was thus found to cover well known
spectral states of this class, namely the very high (VHS), high
(HS), intermediate (IS) and low (LS) states, characterized
respectively by decreasing ratios between the flux of the MCD to the
power-law component of their spectra (see Remillard et al.~2002a; Remillard \& McClintock~2006). At the same time it exhibited
a QPO at frequency $\nu \simeq 5$ Hz along with its first harmonic
and subharmonic frequencies at $\nu \simeq 10~{\rm and}~2.5$ Hz
respectively \cite{Wijnands99}. Such QPOs have been empirically
categorized as type A, B or C, depending on the QPO frequency,
amplitude (rms \%), phase-lag, Q-value and on the photon energy
spectrum. \cite{Remillard02a,Remillard02b}, among others, have
reported the presence of high frequency QPOs (HFQPOs) at frequencies
184 and 276 Hz, consistent with the often seen 2:3 ratio.
Their analysis also indicated the disappearance of the QPO power as
the power-law component of the spectrum decreased below $\sim 20\%$
of the total X-ray flux.

{\em 2. XTE~J1859+226:} This is another GBHC that exhibits rich QPO
behavior including several low frequency QPOs (LFQPOs) in the
frequency range $1-9$ Hz, including the presence of subharmonic
frequencies like XTE~J1550-564 (Casella et al.~2004). These authors
noted also the occurrence of type A, B, C QPOs in this source. This
source, however, exhibits different behavior in the time lags between
the fundamental and the first harmonic from that of XTE~J1550-564
(both of which have hard lags), in that its subharmonic exhibits
soft lag while its fundamental hard.

{\em 3. GRS~1915+105:} This is a GBHC source, known also as a
microquasar because it exhibits VLBI outflows of relativistic
speeds. It is the GBHC source with the highest mass function,
indicating a black hole of mass $M \simeq 15 M_{\odot}$. It is known
to exhibit statistically significant, high-Q, harmonically spaced
LFQPOs, with fundamental frequency $\nu \simeq 67$ mHz and its
first, second, third and possibly even higher harmonics (Cui~1999); it also
exhibits a broad QPO at $\nu \simeq 0.841$ Hz (see
Fig.~\ref{fig:ps}c) and pair of HFQPOs at frequencies $\nu_{1,2} =
40, 67$ Hz (Cui~1999, Strohmayer~2001), having the commensurate
ratio of 2:3, a common property of the HFQPOs of GBHCs.

{\em 4. Cygnus X-2:} This is a typical LMXB Z--source containing a
neutron star accreting at the Eddington limit
(van der Klis~2004). Hasinger et al.~(1986) discovered a QPO at
frequency 30 Hz increasing to 45 Hz with increasing source
luminosity. HFQPOs (at $\nu_l \sim 700$ Hz and $\nu_u\sim 1000$ Hz)
were then discovered by \cite{Wijnands98}, while correlations
between the QPO frequencies, the flux and the spectral index were
established in \cite{TKS07}.

   \begin{figure*}
   \centering
   \includegraphics[trim=0in 0in 0in
0in,keepaspectratio=false,width=2.7in,angle=0]{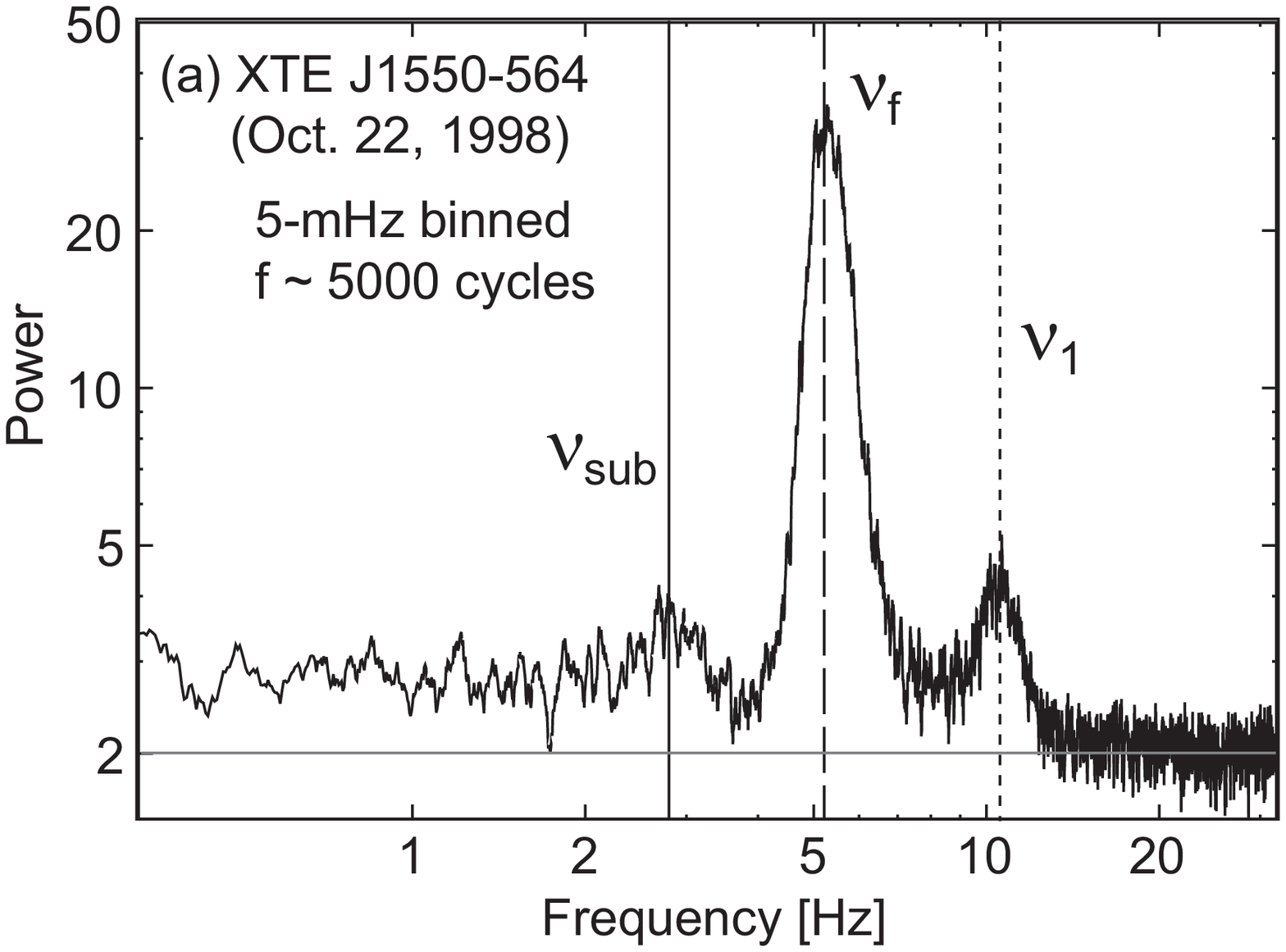}
   \includegraphics[trim=0in 0in 0in
0in,keepaspectratio=false,width=2.7in,angle=0]{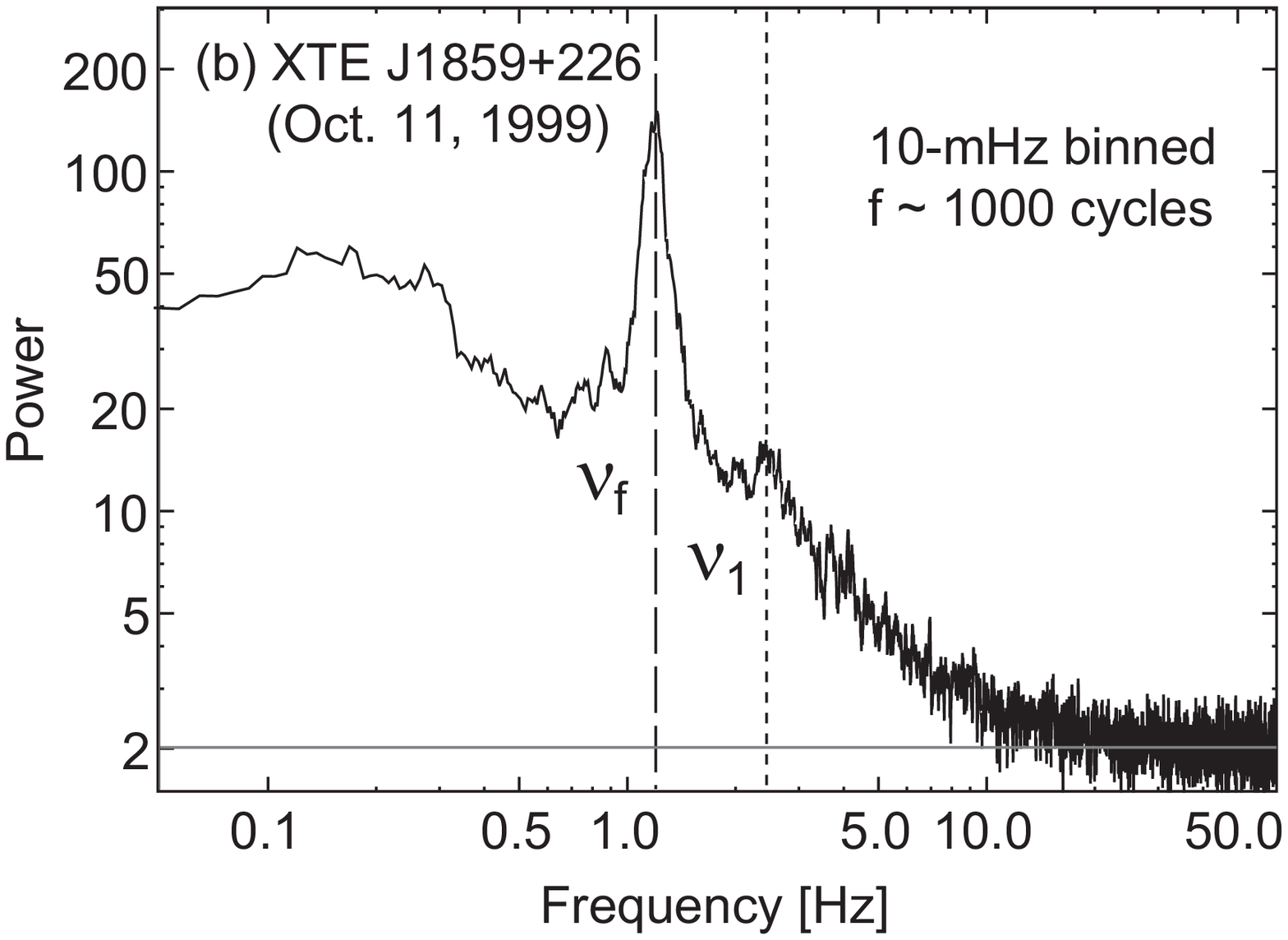}
\\
\includegraphics[trim=0in 0in 0in
0in,keepaspectratio=false,width=2.7in,angle=0]{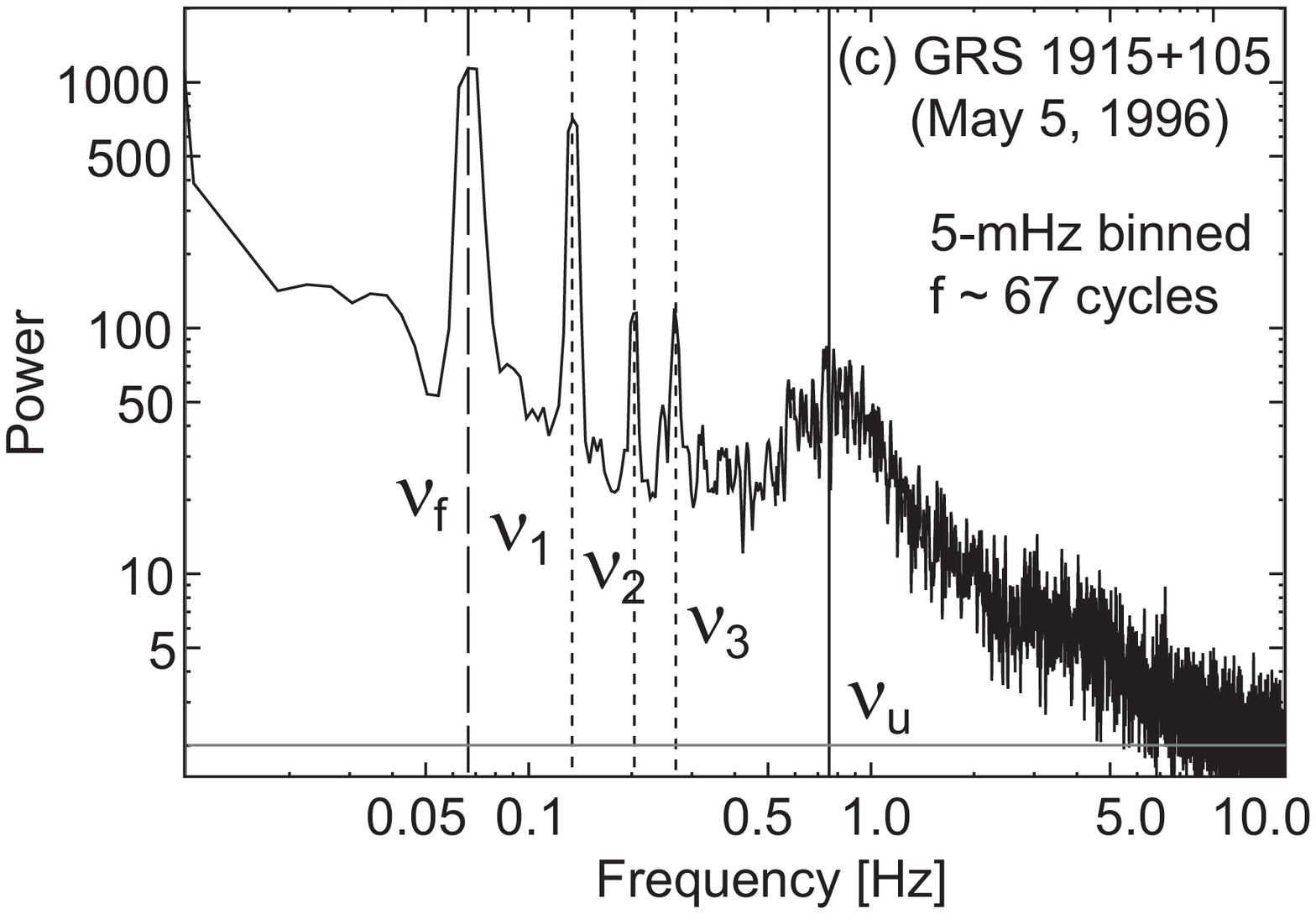}
   \includegraphics[trim=0in 0in 0in
0in,keepaspectratio=false,width=2.7in,angle=0]{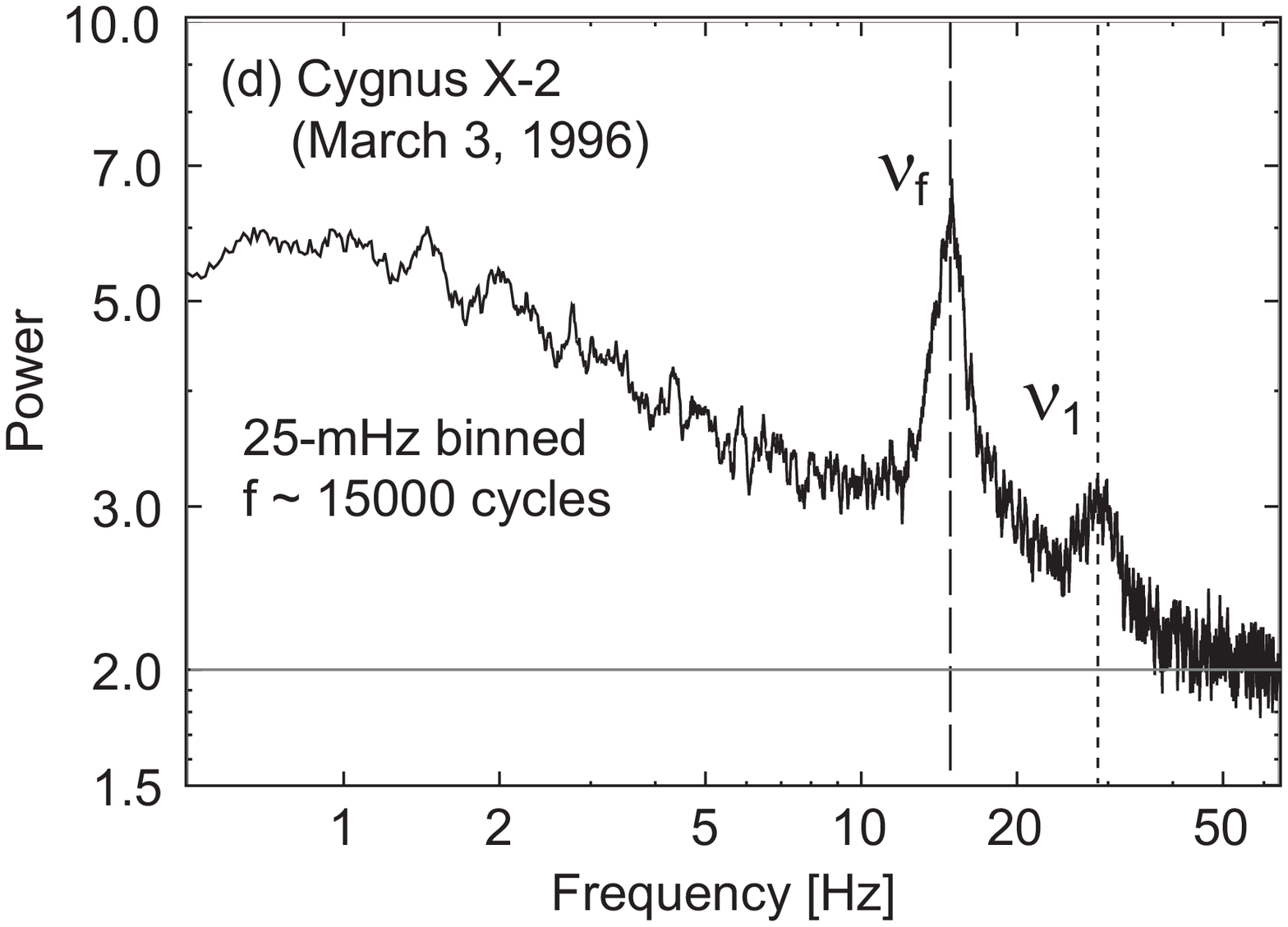}
   \caption{Leahy-normalized power density spectra (PDS) annotating the
identified low-frequency features for (a) XTE~J1550-564 ($\nu_{\rm
sub} \sim 2.8, \nu_f \sim 5.3$ and $\nu_1 \sim 10.9$ Hz), (b)
XTE~J1859+226 ($\nu_f \sim 1.2$ and $\nu_1 \sim 2.4$ Hz), (c)
GRS~1915+105 ($\nu_f \sim 67, \nu_1 \sim 136, \nu_2 \sim 205, \nu_3
\sim 267$ and $\nu_u \sim 781$ mHz) and (d) Cygnus~X-2 ($\nu_f \sim
14.9$ and $\nu_1 \sim 29.1$ Hz) with the spectral resolution and
(fundamental) QPO cycles $f \equiv \nu_f T$ indicated where $T
\backsimeq 1$ ksec (see the text for notations). [{\em See
electronic edition of Journal for a color version of this figure.}]}
              \label{fig:ps}%
    \end{figure*}

   \onlfig{1}{
   \begin{figure*}
   \centering
   \includegraphics[trim=0in 0in 0in
0in,keepaspectratio=false,width=2.7in,angle=0]{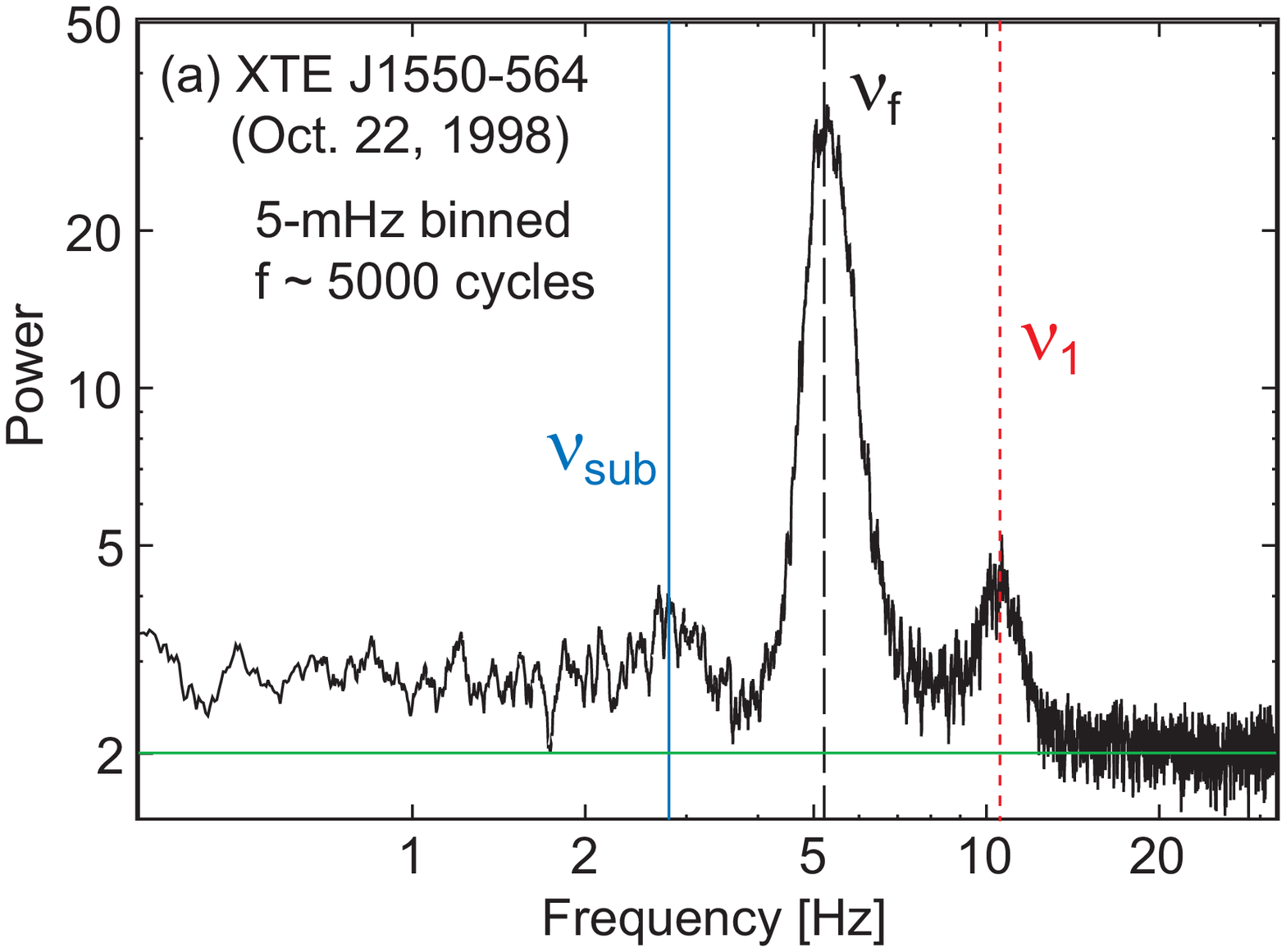}
   \includegraphics[trim=0in 0in 0in
0in,keepaspectratio=false,width=2.7in,angle=0]{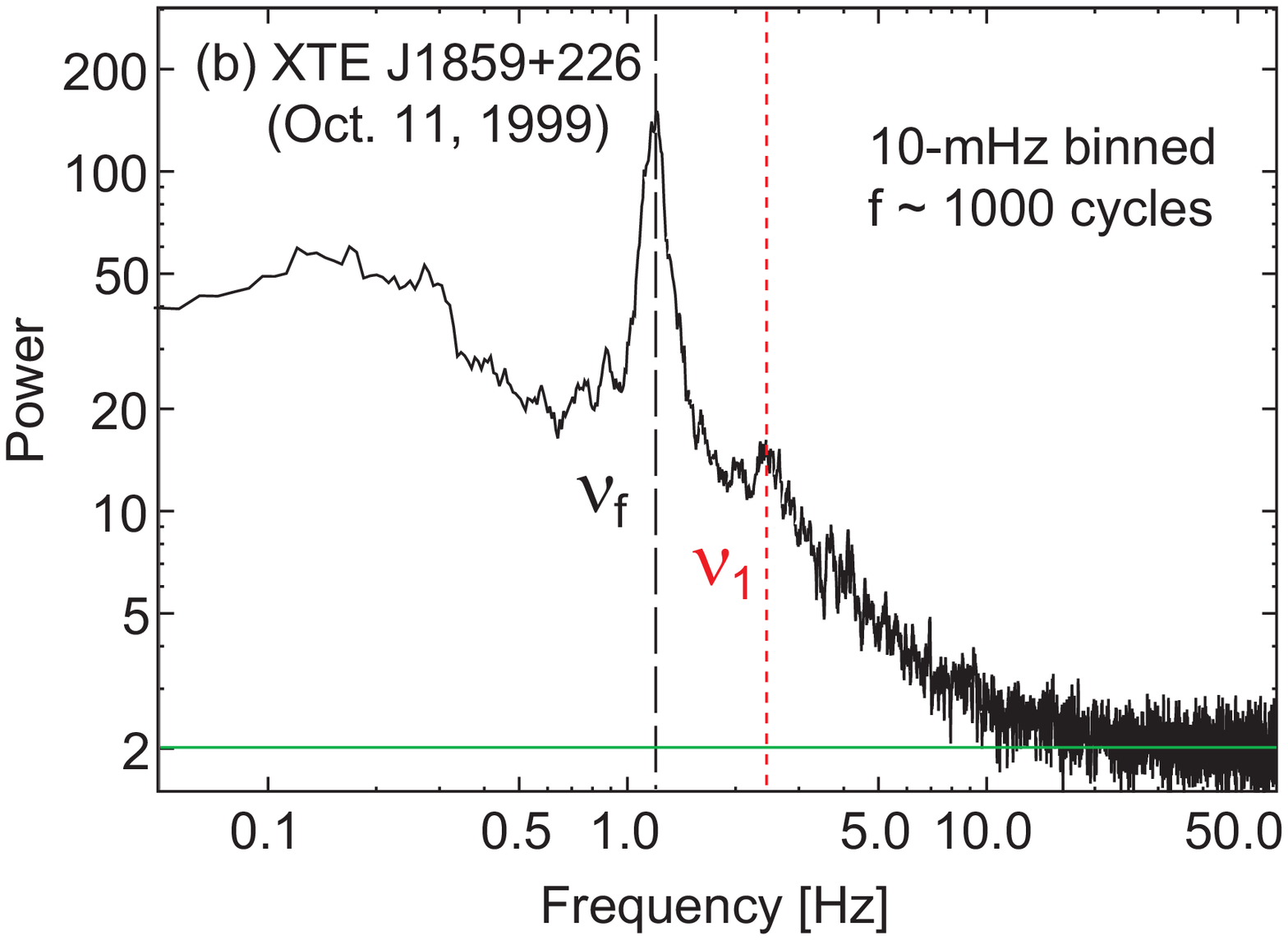}
\\
\includegraphics[trim=0in 0in 0in
0in,keepaspectratio=false,width=2.7in,angle=0]{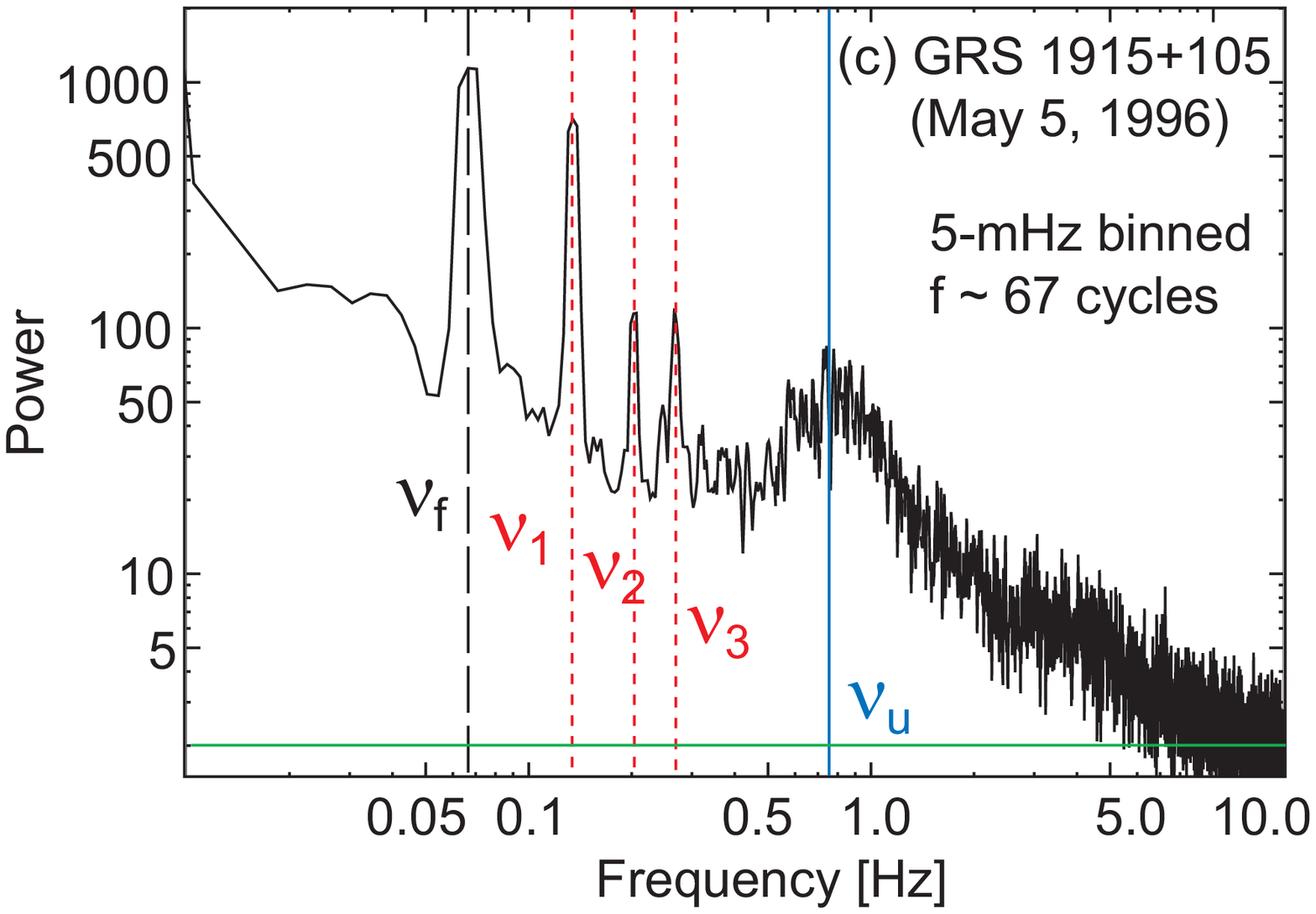}
   \includegraphics[trim=0in 0in 0in
0in,keepaspectratio=false,width=2.7in,angle=0]{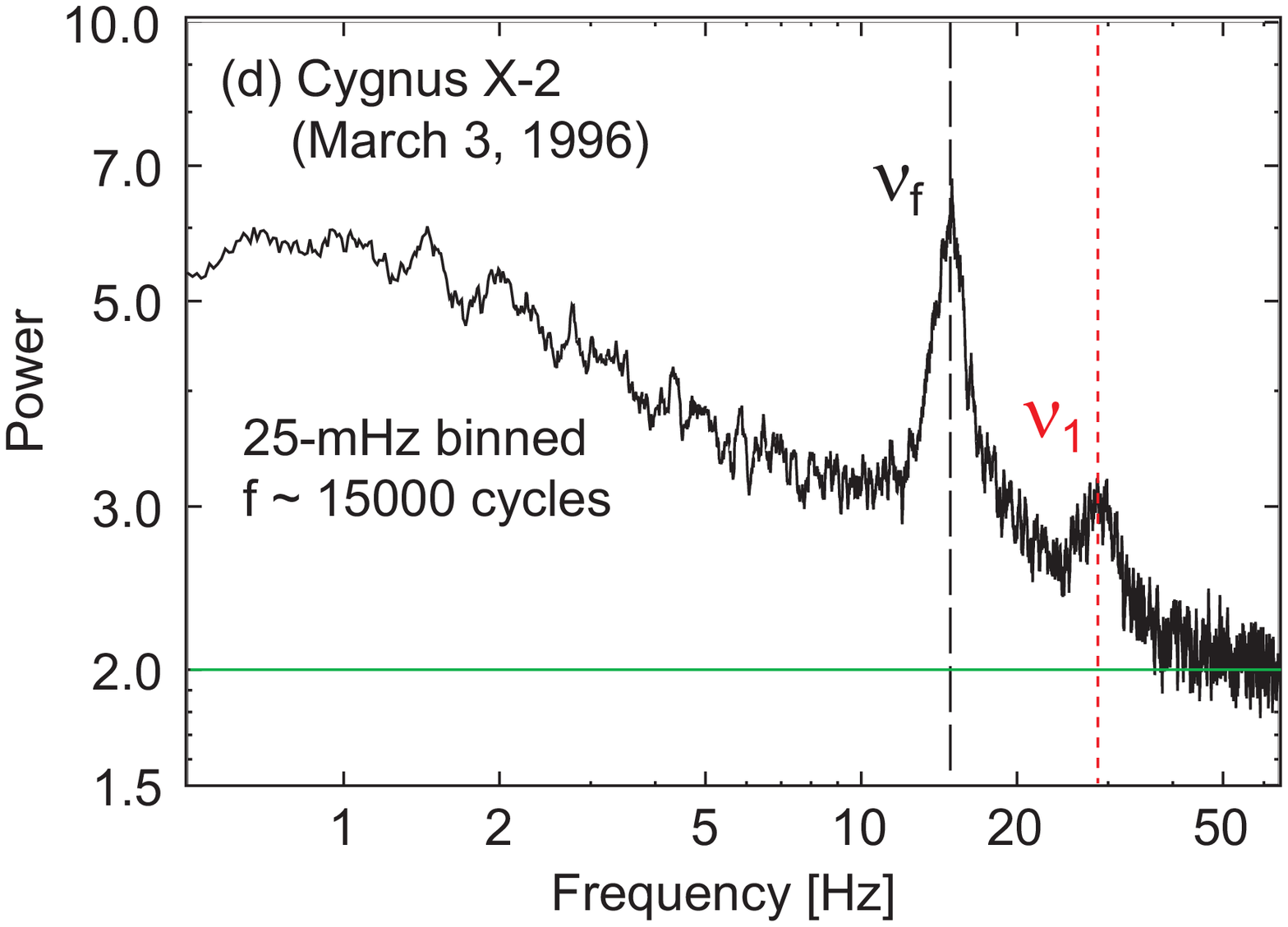}
   \caption{Leahy-normalized power density spectra (PDS) annotating the
identified low-frequency features for (a) XTE~J1550-564 ($\nu_{\rm
sub} \sim 2.8, \nu_f \sim 5.3$ and $\nu_1 \sim 10.9$ Hz), (b)
XTE~J1859+226 ($\nu_f \sim 1.2$ and $\nu_1 \sim 2.4$ Hz), (c)
GRS~1915+105 ($\nu_f \sim 67, \nu_1 \sim 136, \nu_2 \sim 205, \nu_3
\sim 267$ and $\nu_u \sim 781$ mHz) and (d) Cygnus~X-2 ($\nu_f \sim
14.9$ and $\nu_1 \sim 29.1$ Hz) with the spectral resolution and
(fundamental) QPO cycles $f \equiv \nu_f T$ indicated where $T
\backsimeq 1$ ksec (see the text for notations). }
              \label{fig:ps}%
    \end{figure*}
    }

In Figure~\ref{fig:ps} we show the Leahy normalized PDS (Leahy
normalization provides the fractional amplitude relative to a
$\chi^2$ process with 2 degrees of freedom which is known to have
amplitude 2 and variance 4; see Leahy et al.~1983) of each of the above sources obtained at
the specific date indicated along with the binned frequency
resolution and the number of QPO cycles $f\equiv \nu_f T$ at the
fundamental QPO frequency $\nu_f$ where $T$ is the duration of the
observed LC. While the variability properties are not stationary but
vary with (or without) the source flux, we believe that the above
are representative of the general PDS form of accretion powered
sources; these exhibit a flat (white noise-type) or red noise low
frequency regime, complemented in some cases by LFQPOs, which breaks
to a steeper spectrum above a few Hz. At higher frequencies the PDS
of the GBHCs continue to decrease (with exception the presence of
HFQPOs), while those of neutron stars recover at the highest
frequencies (kHz) where they exhibit prominent kHz QPOs. This
difference between the overall form of the PDS of GBHCs and LMXB
neutron stars has, in fact, been proposed as a discriminant of the
nature of the accreting object (Sunyaev \& Revnivtsev~2000). These
PDS while not identical they all exhibit prominent QPOs with at
least one harmonic beyond the fundamental, the latter being their
unifying characteristic. In fact, the most similar in their overall
PDS form are the GBHC XTE~J1859+226 and the neutron star Cyg X-2.

The data used in producing the ACFs and PDSs in this paper contain
photons in the energy range $5 \lsim E \lsim 13$ keV for all
sources; they were also obtained at the following dates: (a)
XTE~J1550-564 in Oct. 22 (1998), (b) XTE~J1859+226 in Oct.~11
(1999), (c) GRS~1915+105 in May 5 (1996), and (d) Cygnus~X-2 in
March 3 (1996), with exposure times of approximately $T \simeq 1$
ksec in all cases.

\section{Autocorrelation Function (ACF) Analysis}

For a discrete LC, $I(t_i)$,(photon counts) defined at times $ t_i$,
the mean-subtracted normalized ACF is given by
\begin{eqnarray}
\textmd{ACF}(\tau_i) &\equiv&  \sum_{k=1}^N
\left[\bar{I}(t_k)  \bar{I}(t_k+\tau_i)\right] / \sum_{k=1}^N
\bar{I}^2(t_k) \ , \label{eq:ACF}
\end{eqnarray}
where $N$ is the entire number of time bins in the LC (sampling
number) with a constant sampling time (temporal resolution) of
$\Delta T \equiv t_{i+1}-t_i$ where $T = N \Delta T$ with $\tau_i$
denoting the characteristic time-lag in the LC (see, for some of
its applications, Fukumura \& Kazanas~2008, Fukumura, Kazanas, \&
Stephenson~2009); $\bar I(t_k)$ is the mean-subtracted LC defined as
\begin{eqnarray}
\bar{I}(t_k) \equiv I(t_k)-I_0 \ , \label{eq:I}
\end{eqnarray}
with $I_0$ the mean value of $I(t_k)$ over the interval $T$, a
definition which allows the ACF($\tau_i)$ to be either positive or
negative. Note that with this definition the ACF obtains its maximum
value of unity at $\tau_i=0$. Because of the finite length of the
observed signal (i.e. the LC) the ACF must also, by definition,
decline to zero at $\tau_i=T$. 

   \begin{figure*}
   \centering
   \includegraphics[trim=0in 0in 0in
0in,keepaspectratio=false,width=2.7in,angle=0]{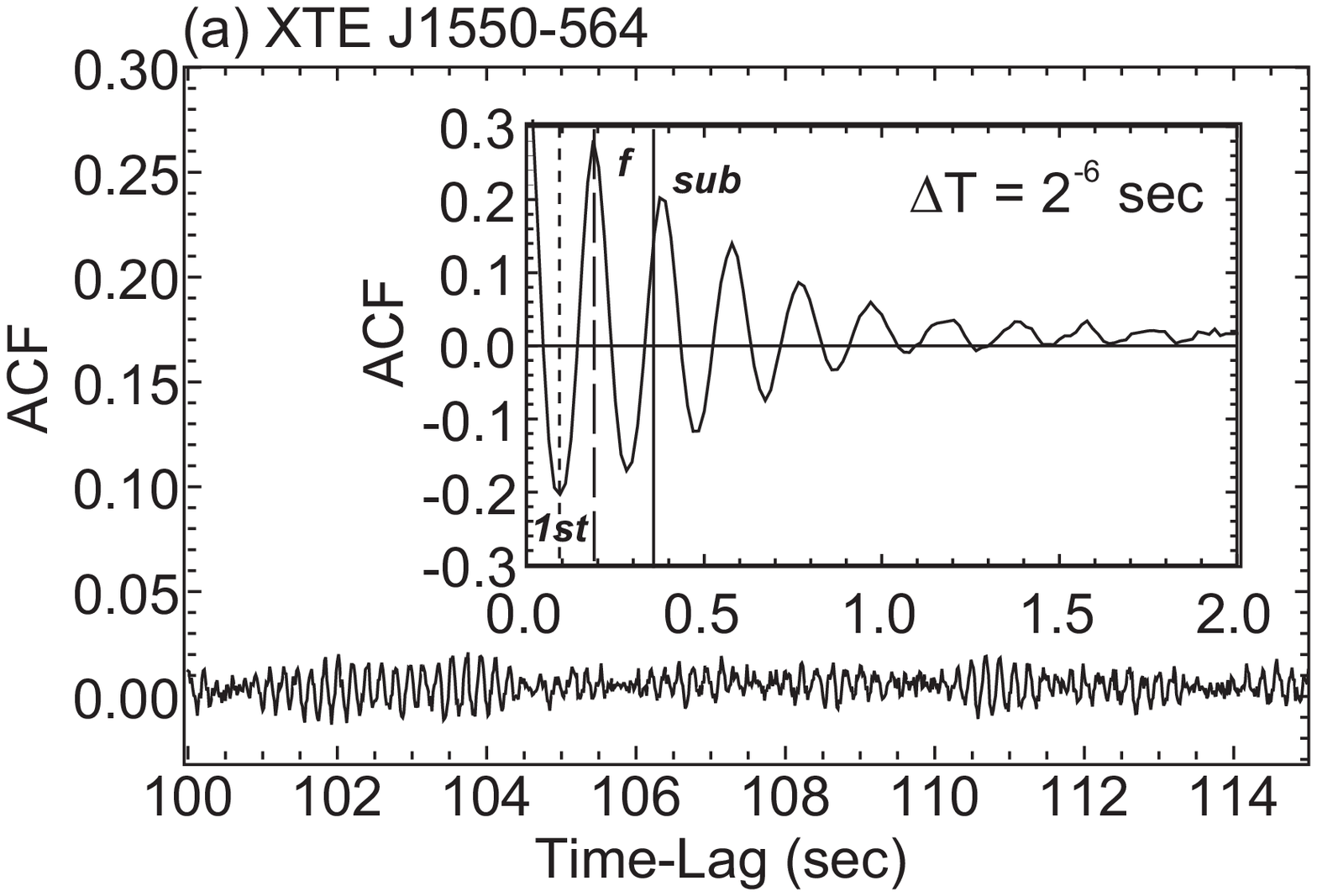}
   \includegraphics[trim=0in 0in 0in
0in,keepaspectratio=false,width=2.7in,angle=0]{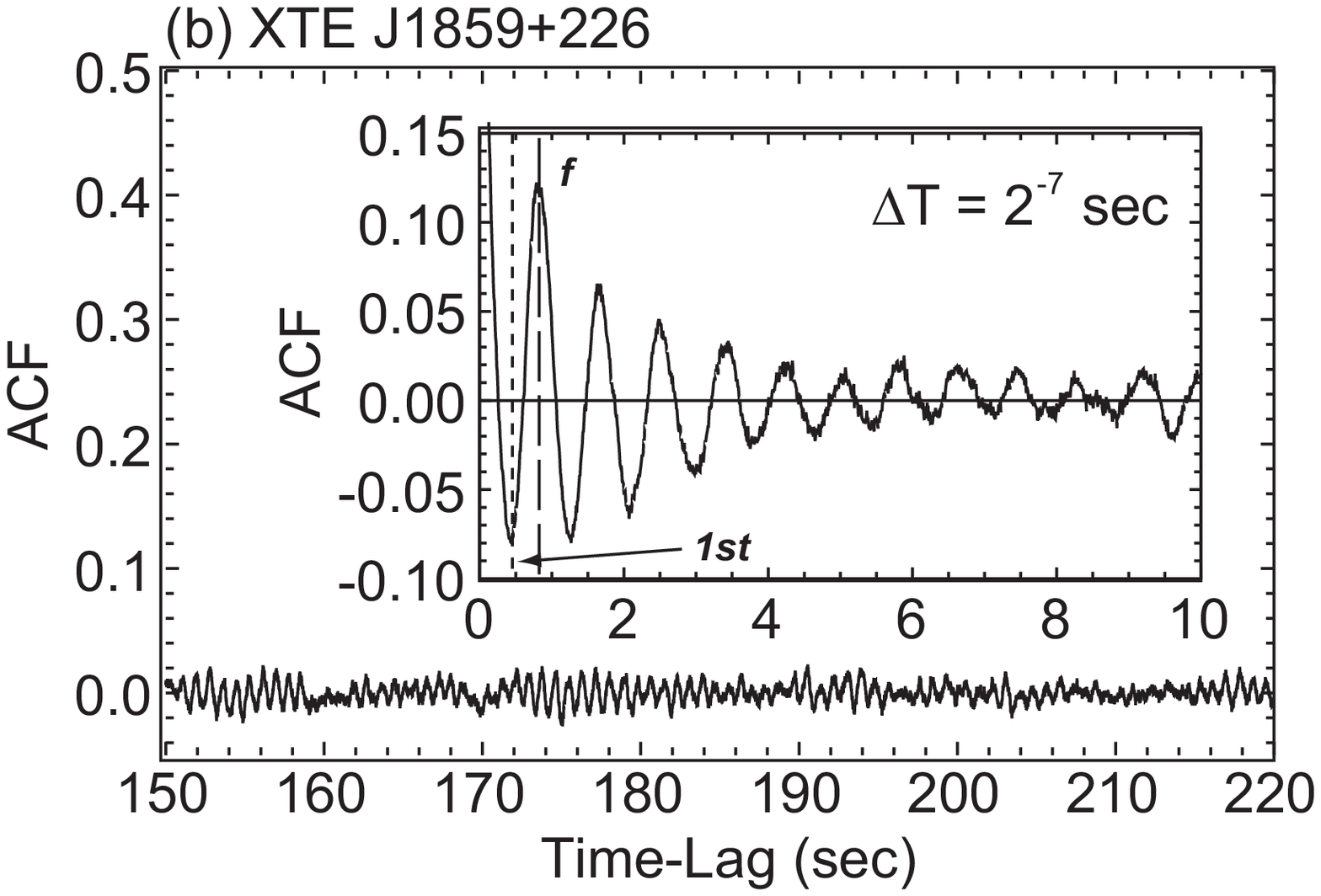}
\\
\includegraphics[trim=0in 0in 0in
0in,keepaspectratio=false,width=2.7in,angle=0]{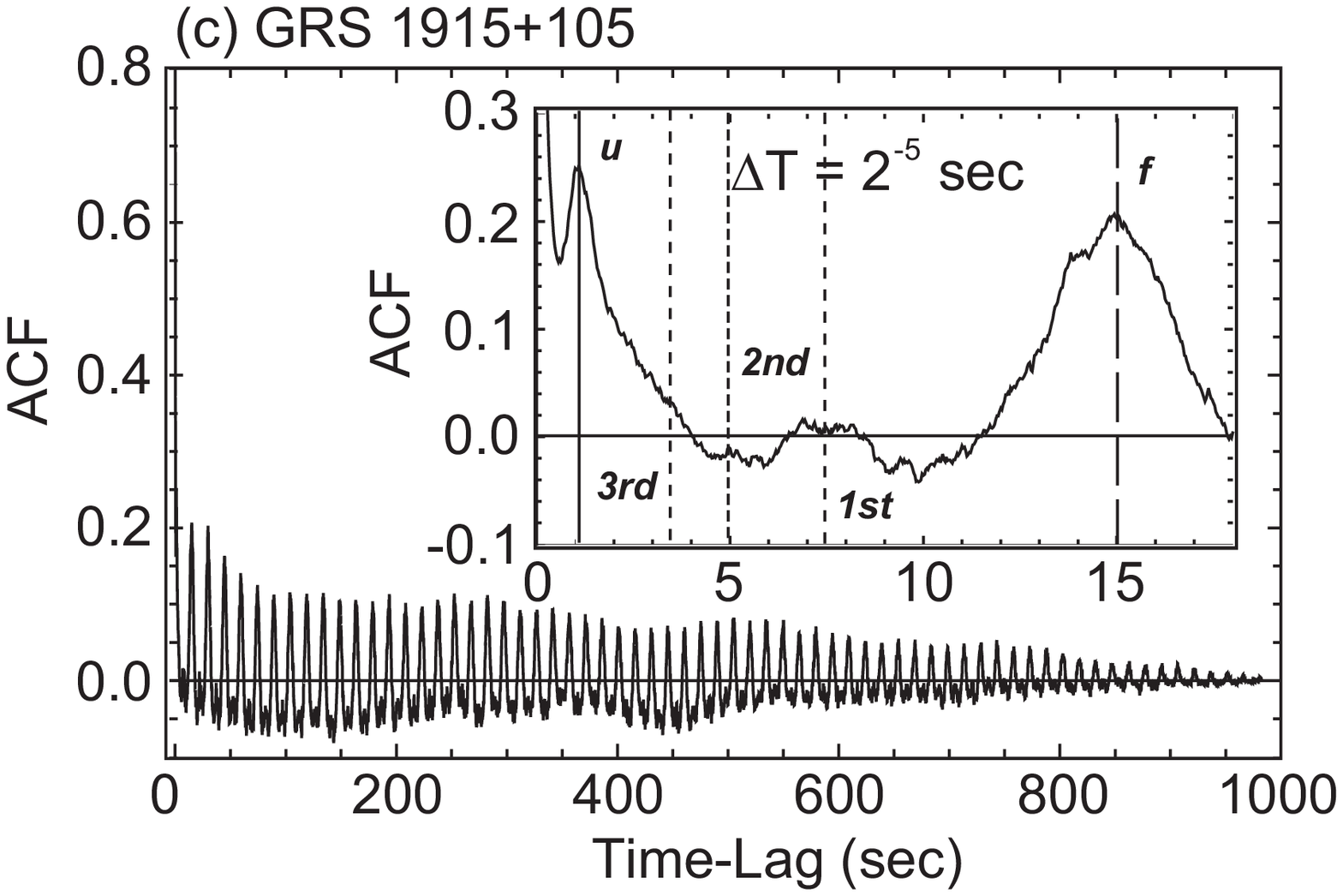}
   \includegraphics[trim=0in 0in 0in
0in,keepaspectratio=false,width=2.7in,angle=0]{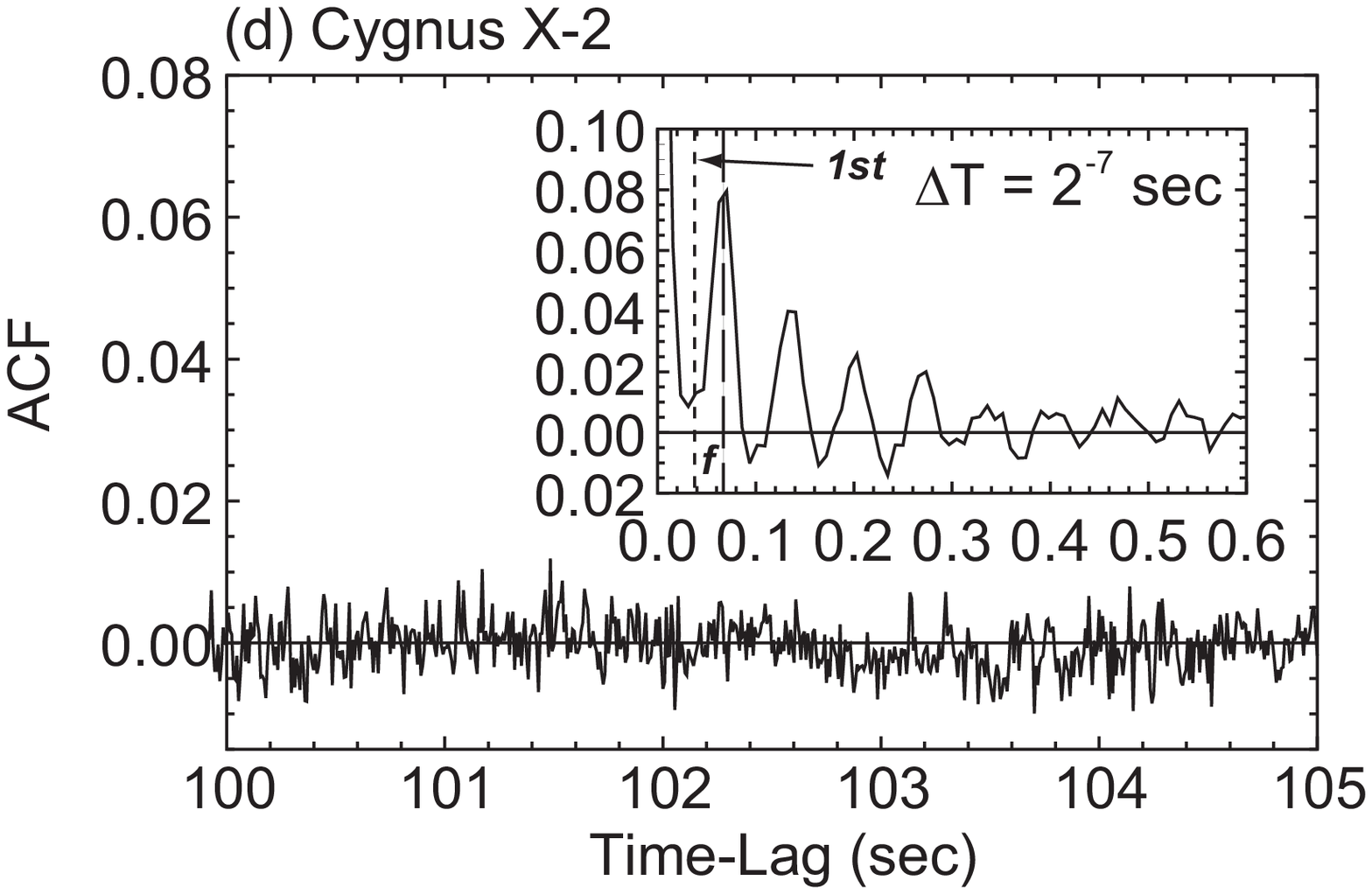}
   \caption{ACFs corresponding to Figure~\ref{fig:ps} for both long
time-lags (extended to hundreds of seconds) and shorter time-lags
(insets) with the temporal resolution $\Delta T$ indicated ($T
\backsimeq 1$ ksec). Vertical lines correspond to the QPO peaks in
the PDS from Figure~\ref{fig:ps}. The error bars in each of the
above ACF has been estimated to be on the order of $\simeq
0.001-0.01$. [{\em See electronic edition of Journal for a color
version of this figure.}]}
              \label{fig:acf}%
    \end{figure*}

   \onlfig{2}{
   \begin{figure*}
   \centering
   \includegraphics[trim=0in 0in 0in
0in,keepaspectratio=false,width=2.7in,angle=0]{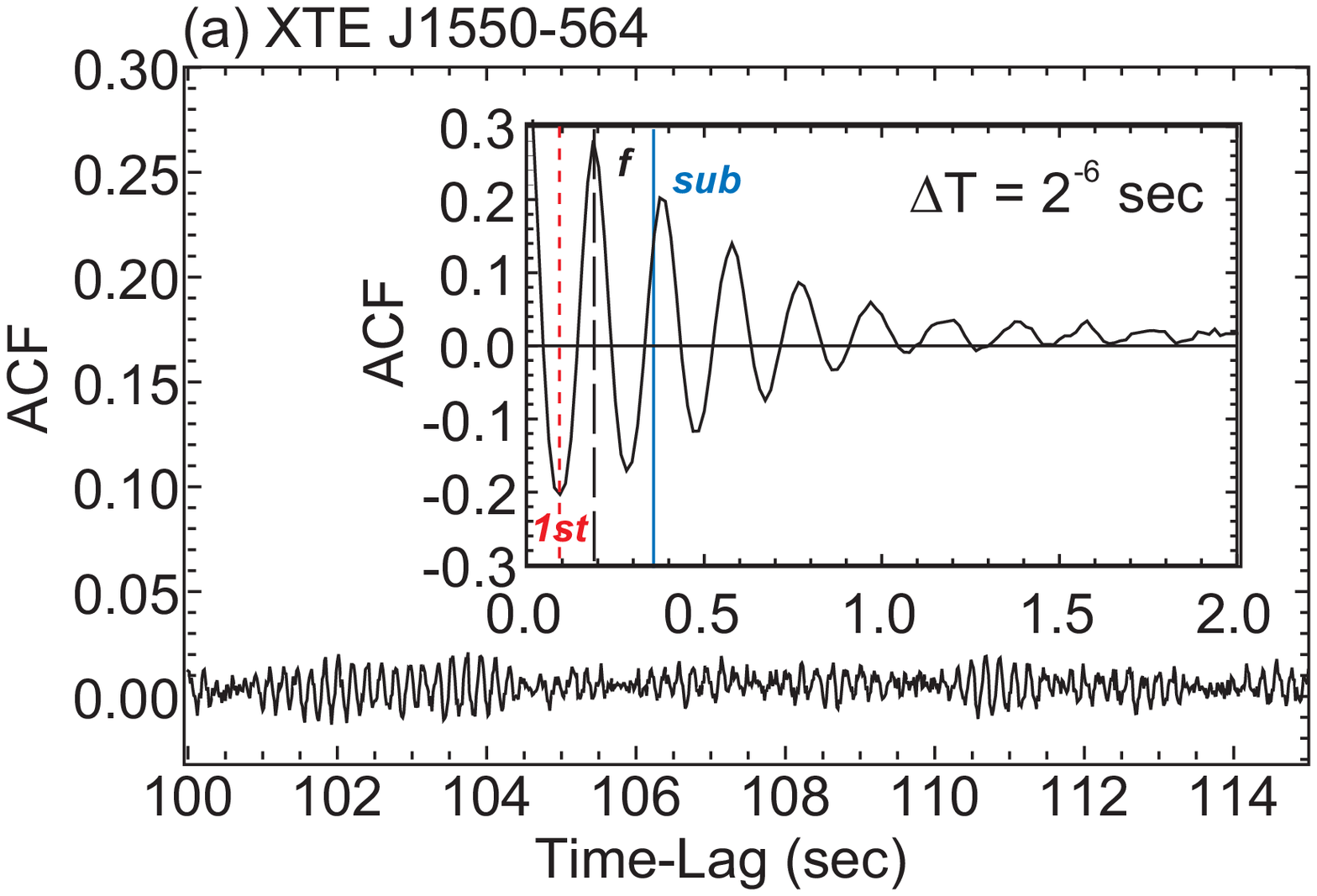}
   \includegraphics[trim=0in 0in 0in
0in,keepaspectratio=false,width=2.7in,angle=0]{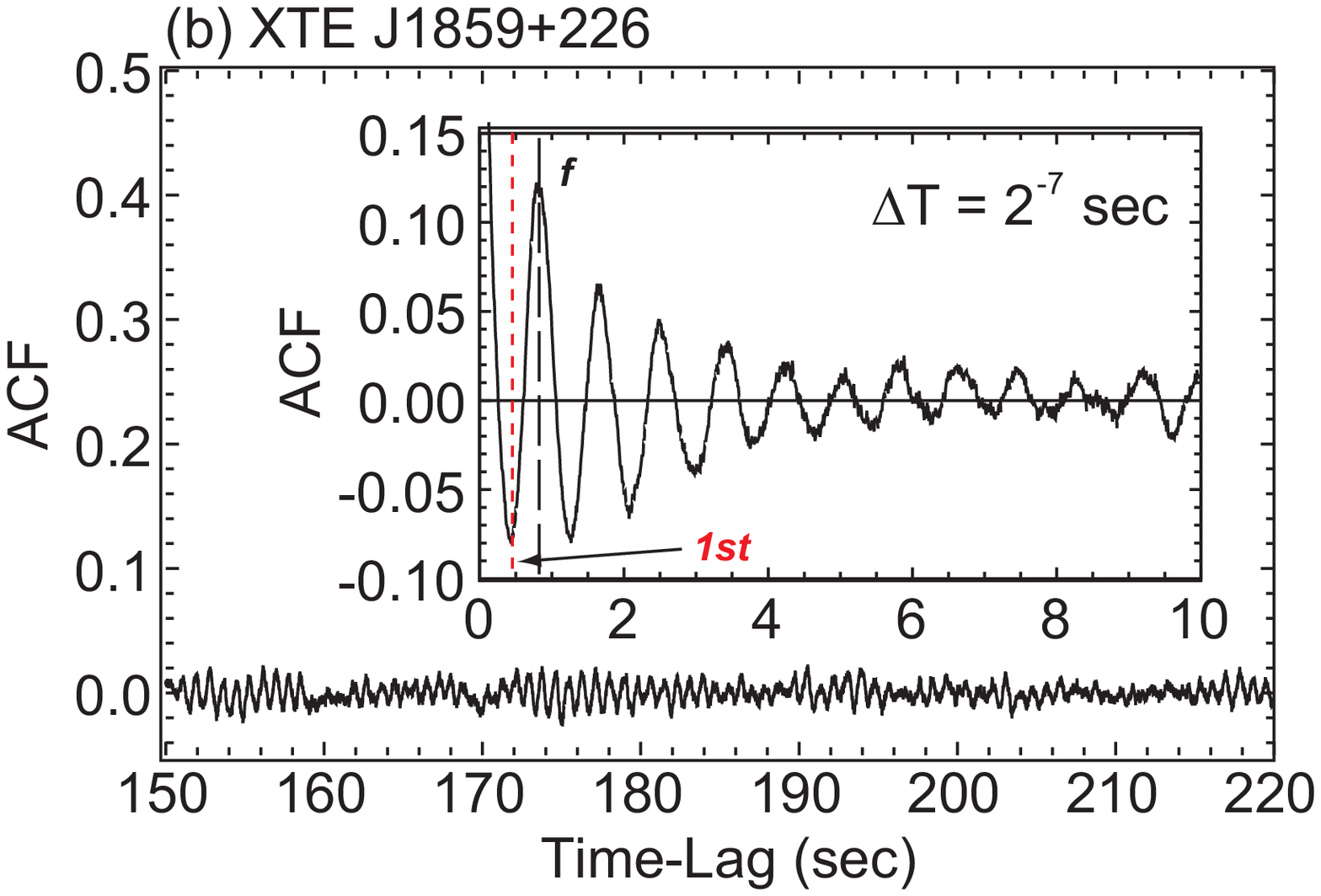}
\\
\includegraphics[trim=0in 0in 0in
0in,keepaspectratio=false,width=2.7in,angle=0]{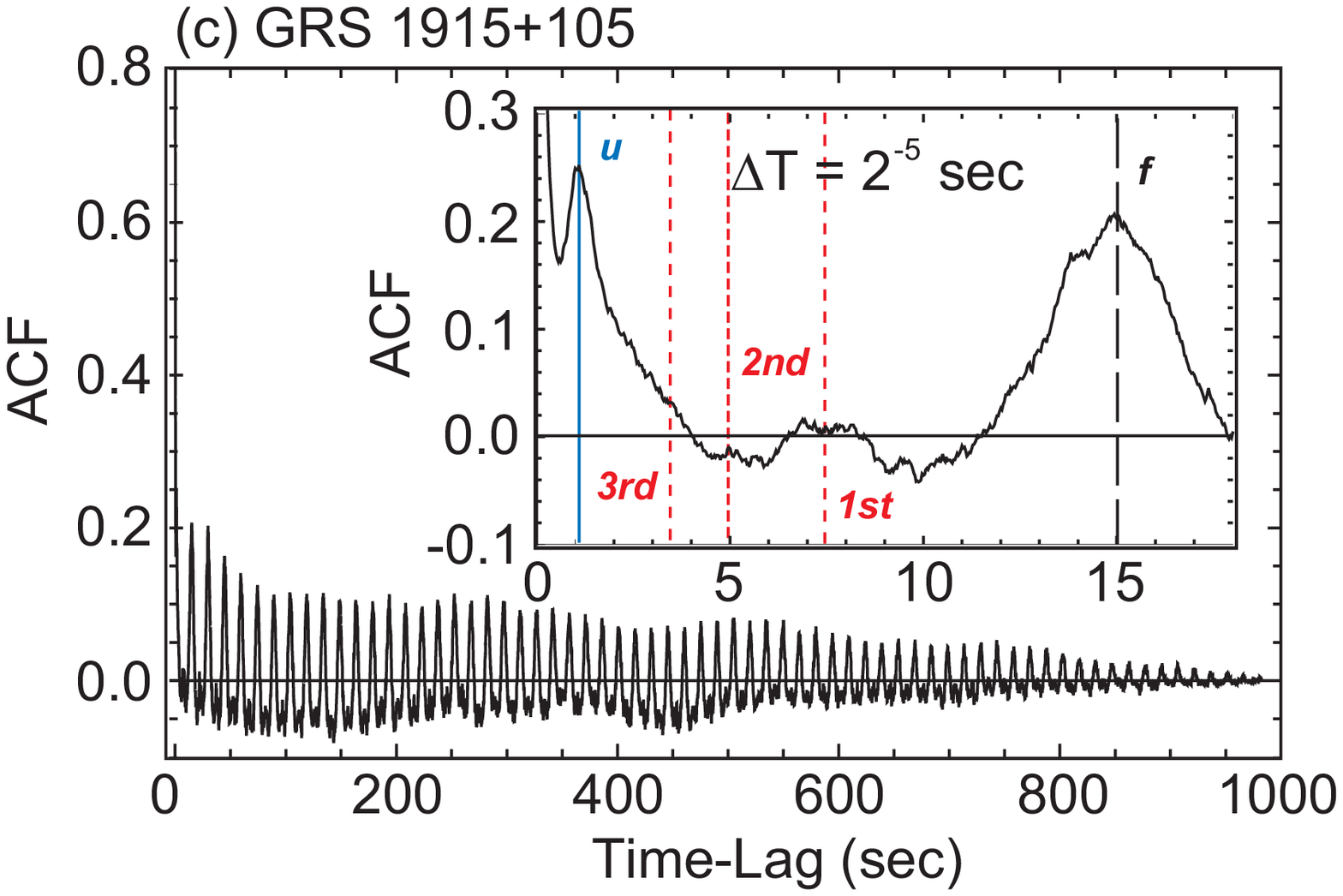}
   \includegraphics[trim=0in 0in 0in
0in,keepaspectratio=false,width=2.7in,angle=0]{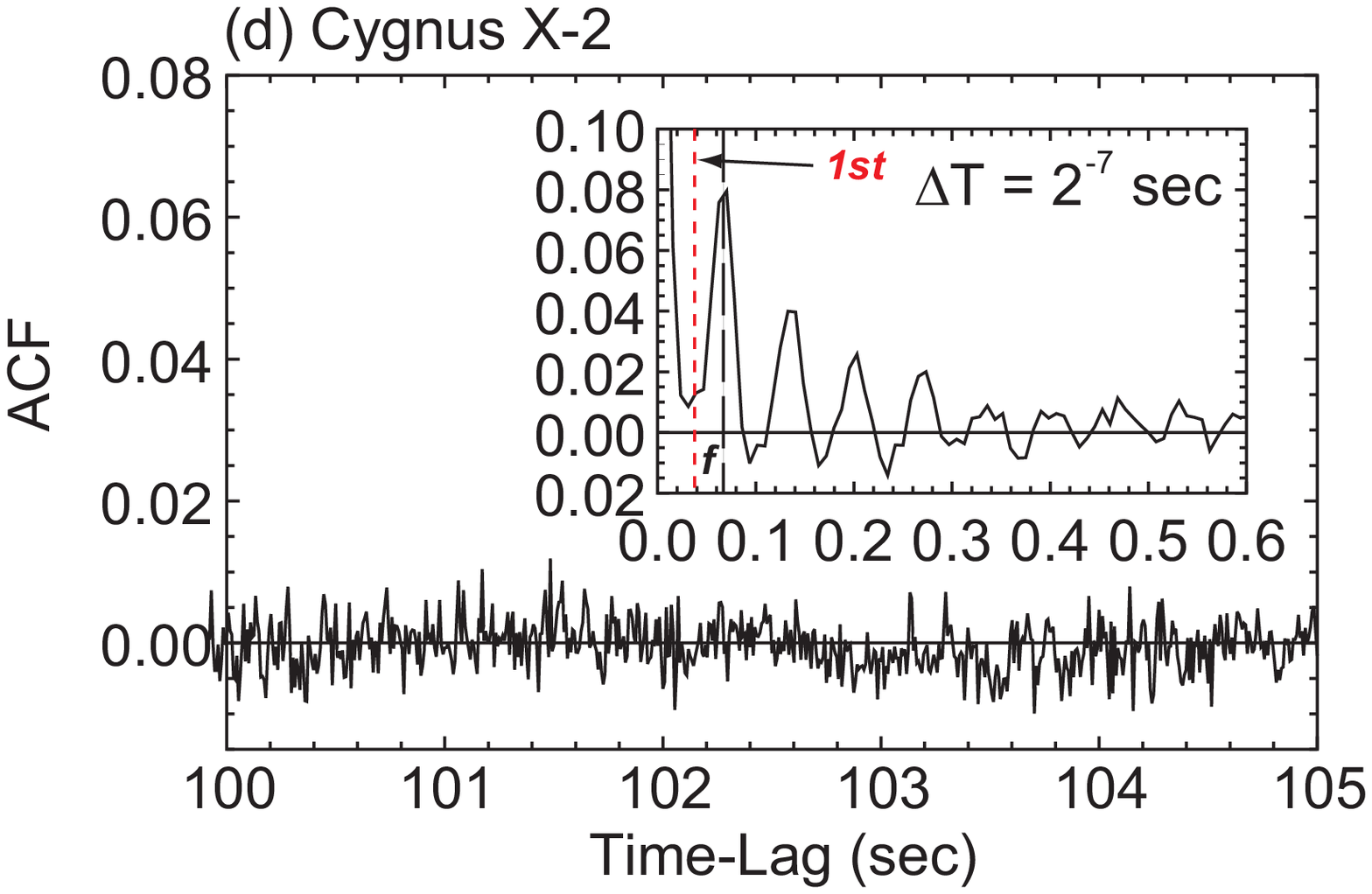}
   \caption{ACFs corresponding to Figure~\ref{fig:ps} for both long
time-lags (extended to hundreds of seconds) and shorter time-lags
(insets) with the temporal resolution $\Delta T$ indicated ($T
\backsimeq 1$ ksec). Vertical lines correspond to the QPO peaks in
the PDS from Figure~\ref{fig:ps}. The error bars in each of the
above ACF has been estimated to be on the order of $\simeq
0.001-0.01$. }
              \label{fig:acf}%
    \end{figure*}
    }
%

%
In Figure~\ref{fig:acf} we show the ACFs corresponding to the LCs
whose PDS are shown in Figure~\ref{fig:ps}. The main figure of each
panel exhibits the ACF at long lags ($\tau \gg 1/\Delta \nu$) with
its behavior near $\tau \simeq 0$ shown in the inset. The
differences and similarities of the ACFs are apparent in these
figures: 1. They all decrease from the value 1 at $\tau =0$ to 0.1
or 0.2 after one oscillation period. 2. The ACF of XTE 1550-564, XTE
1859+226 and Cyg X-2 are quite similar amongst themselves in that
they all exhibit near $\tau = 0$ the form of a damped oscillation
with period that of the QPO fundamental and decay time equal to the
inverse of its width, i.e.  $\tau \simeq 1/\Delta \nu$. 3. The ACF
of GRS 1915+105 is markedly different from those of the other three
sources in that in reaching the value 0.2 after the first period  it
oscillates throughout the duration of the LC at the frequency of the
QPO fundamental and with an amplitude that decreases {\em linearly
rather than exponentially} to zero at lag equal to the observation
time. As such it resembles the long time ACF behavior of a pure
sinusoid or any undamped oscillation with coherent phase.
4. Despite their overall similarity, the ACF of the GBHCs XTE
1550-564 XTE 1859+226 and that of the neutron star Cyg X-2 differ in
their long time behavior:  For $\tau \gg 1/\Delta \nu$ they all
attain small amplitudes $\sim 2 \cdot 10^{-2}$ for the GBHCs in (a)
and (b) and $\lsim 10^{-2}$ for Cyg X-2 in (d); however the GBHC
ACFs exhibit an oscillatory behavior at the QPO fundamental
frequency for lags $\tau$ equal to the observation duration,  while
the about zero fluctuations of the Cyg X-2 ACF appear totally
random.

\section{The Power Spectrum - Autocorrelation Synergy}

\begin{figure*}
   \centering
   \includegraphics[trim=0in 0in 0in
0in,keepaspectratio=false,width=2.7in,angle=0]{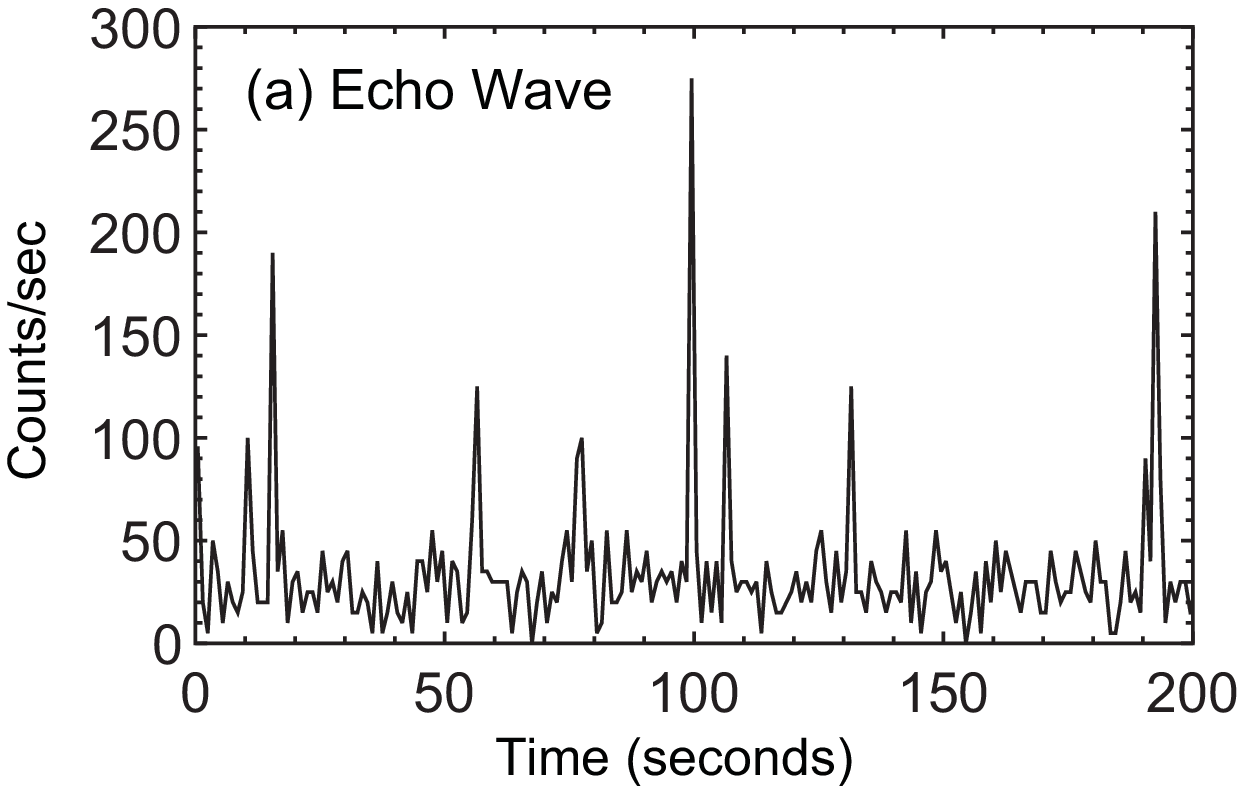}
   \includegraphics[trim=0in 0in 0in
0in,keepaspectratio=false,width=2.7in,angle=0]{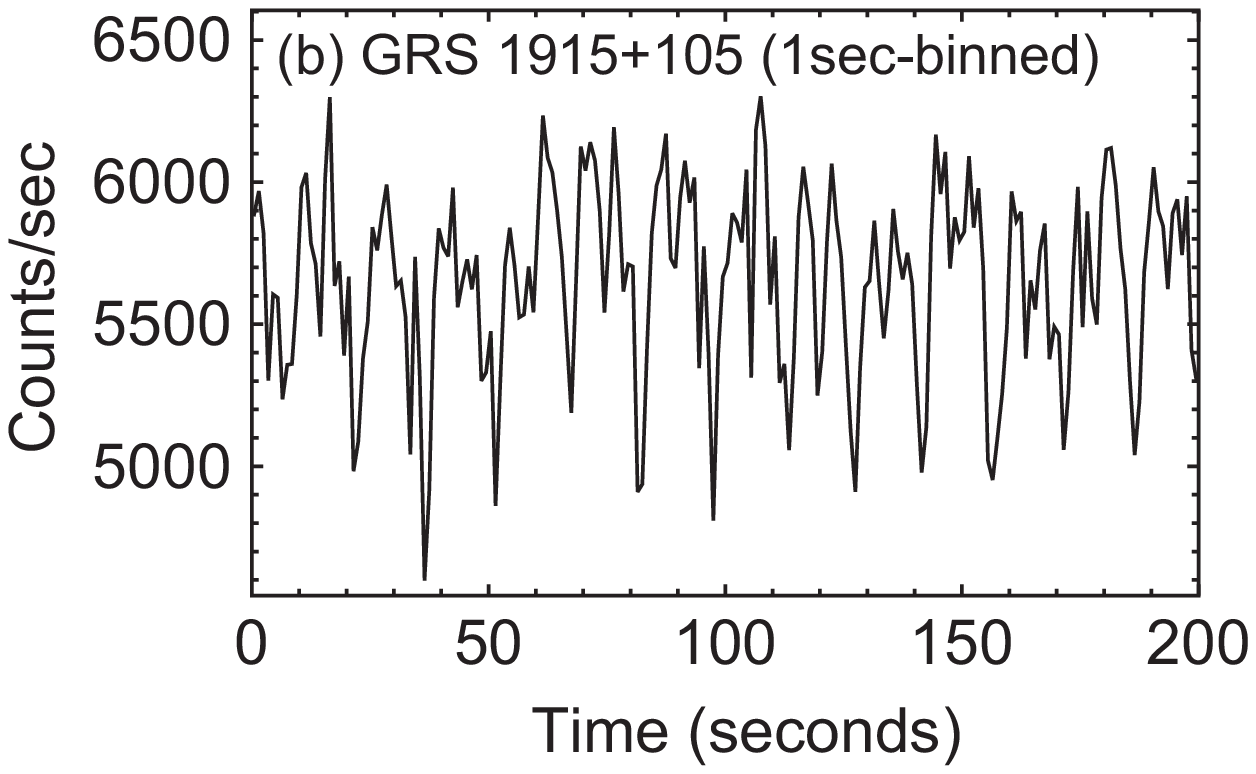}
\\
\includegraphics[trim=0in 0in 0in
0in,keepaspectratio=false,width=2.7in,angle=0]{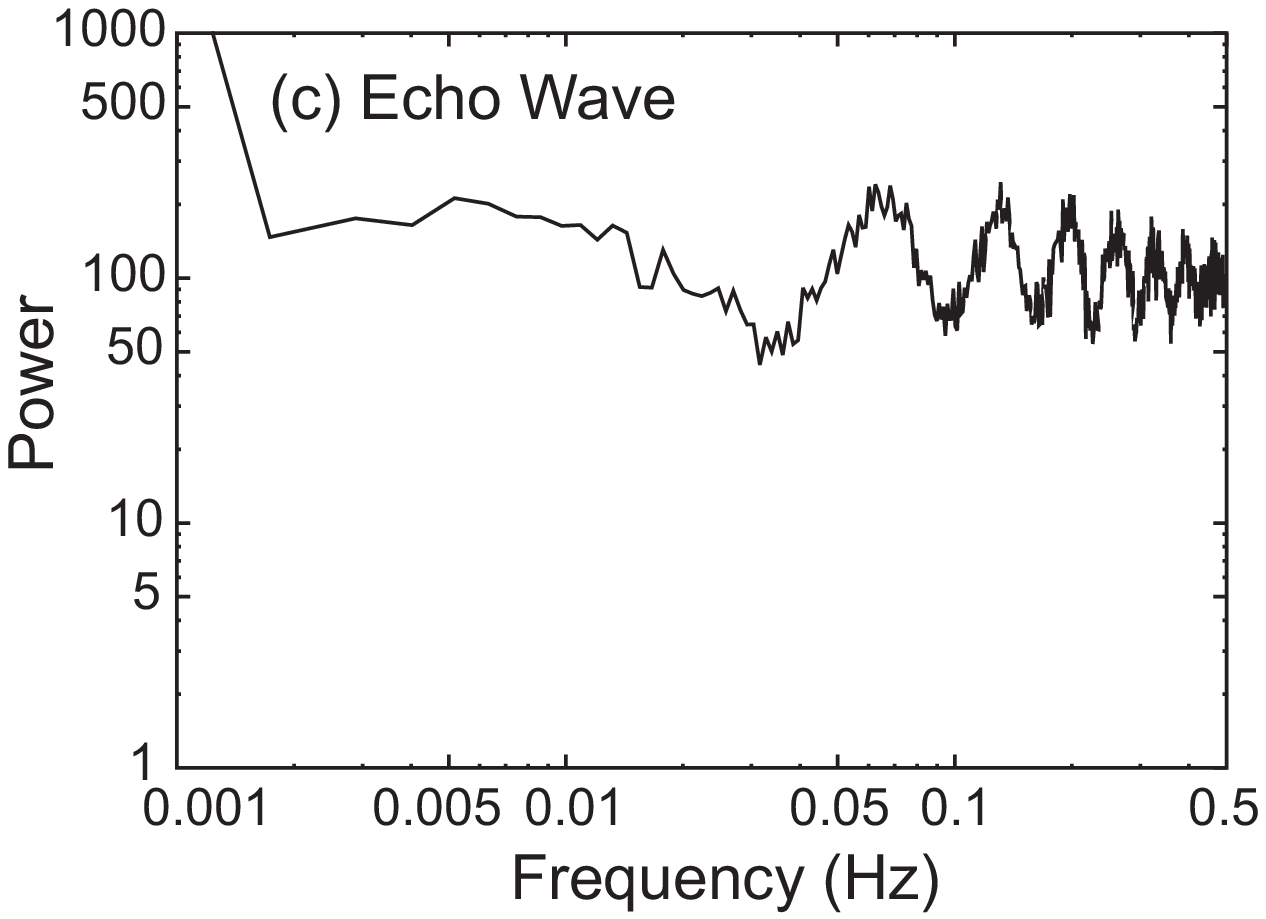}
   \includegraphics[trim=0in 0in 0in
0in,keepaspectratio=false,width=2.7in,angle=0]{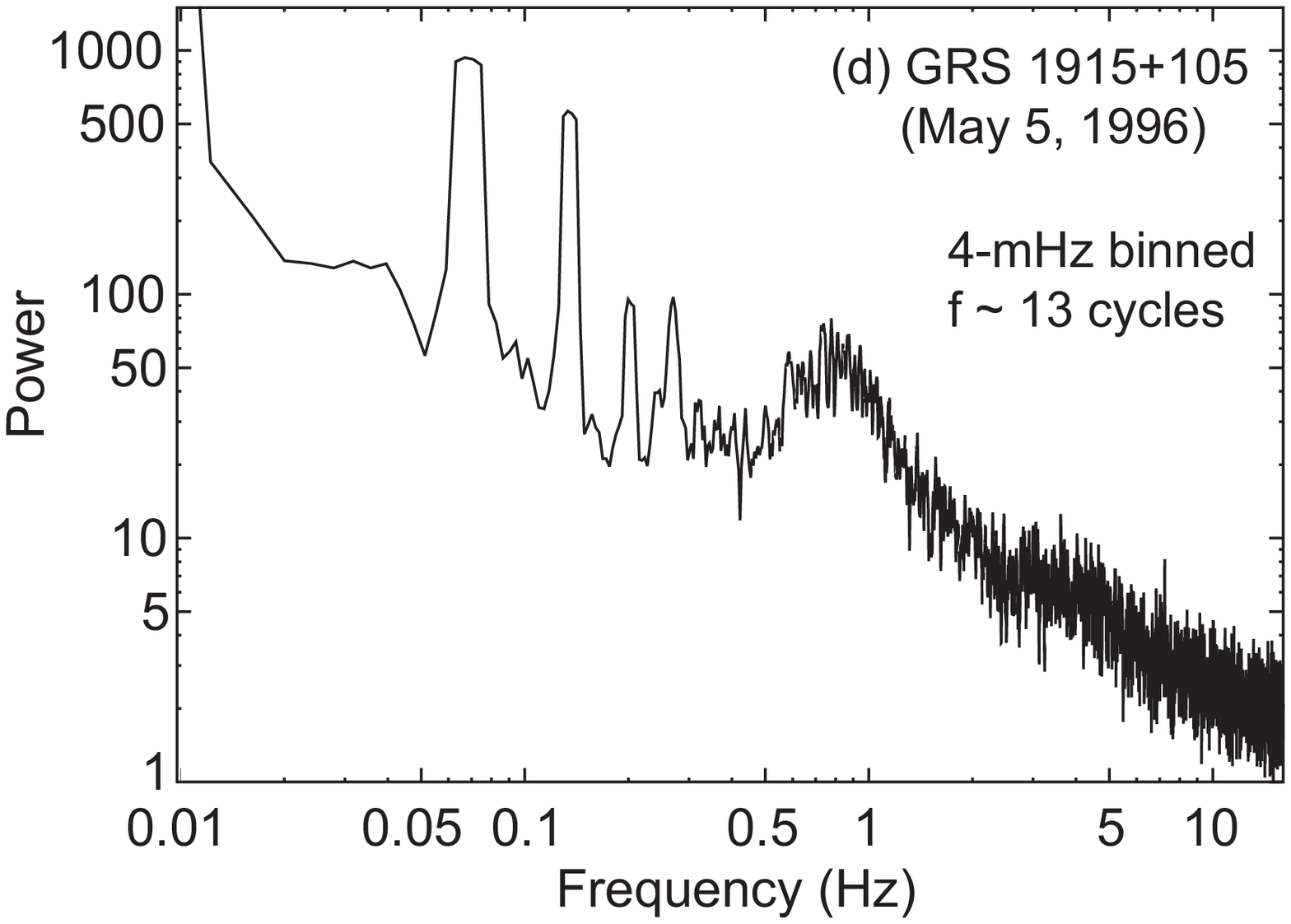}
   \caption{The light curves (upper panels) of $T=200$ sec and the corresponding power
spectra (lower panels) from a Kerr ``echo" as discussed in the text
(left panels) and from observations of GRS~1915+105 (right panels).}
              \label{fig:f3}%
    \end{figure*}

To demonstrate {\it the complementary nature of the PDS and ACF} we
present in this section an explicit example of the synergy between
analysis in the frequency and time domains involving two different
time series which both exhibit well defined harmonically spaced QPOs
in their PDS. The LCs are given in Figure~\ref{fig:f3}a and b, one
of which is a simulated Poisson noise-like LC while the second is
that of GRS 1915+105.
%
%
Visual inspection of these time series suggests the first one
(Fig.~\ref{fig:f3}a) to be totally {\it aperiodic} (and indeed it
was constructed as such) with the second one (Fig.~\ref{fig:f3}b)
exhibiting a roughly {\it periodic} variability with period of order
of 15 sec.

%

   \begin{figure*}
   \centering
   \includegraphics[trim=0in 0in 0in
0in,keepaspectratio=false,width=2.7in,angle=0]{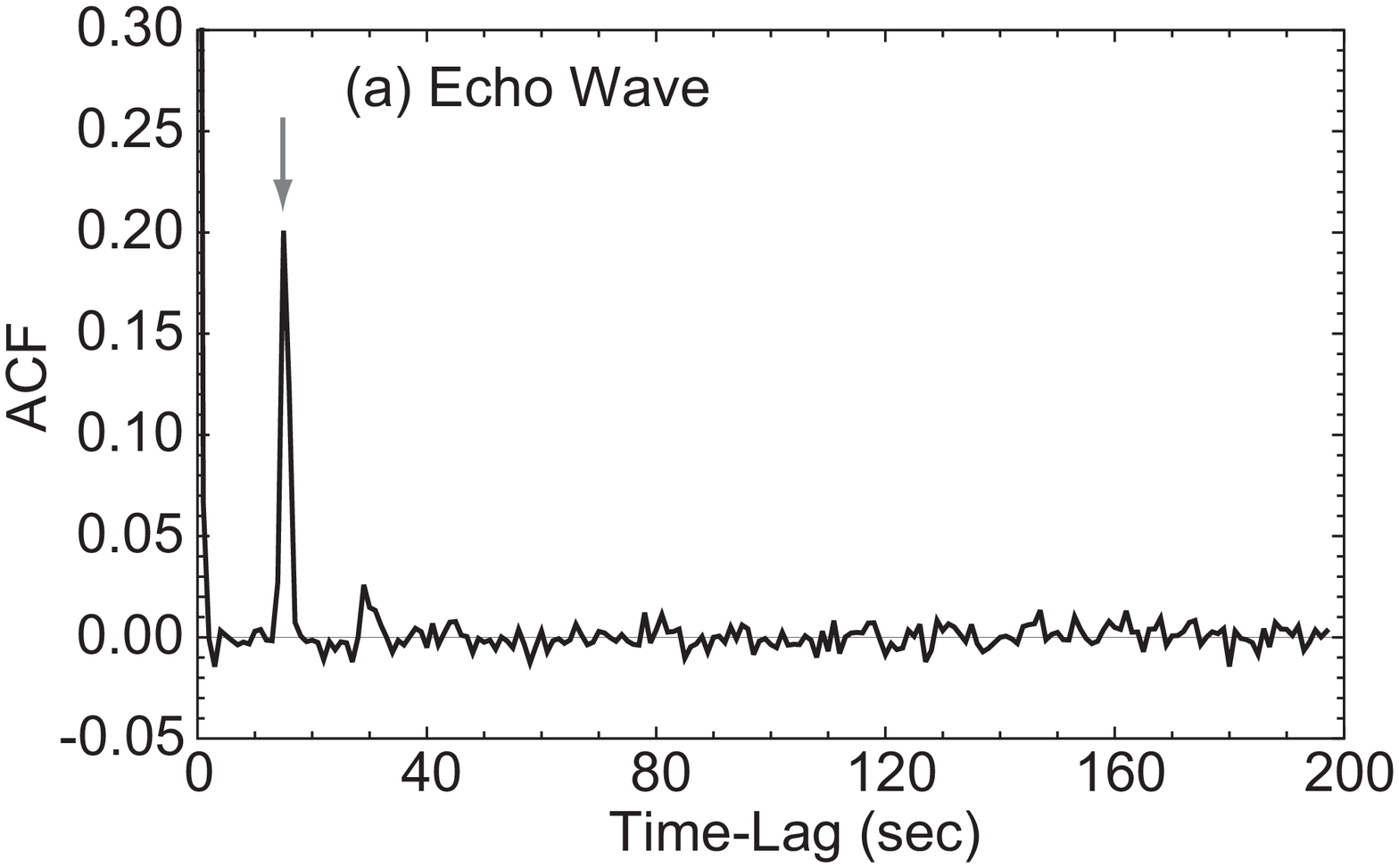}
   \includegraphics[trim=0in 0in 0in
0in,keepaspectratio=false,width=2.7in,angle=0]{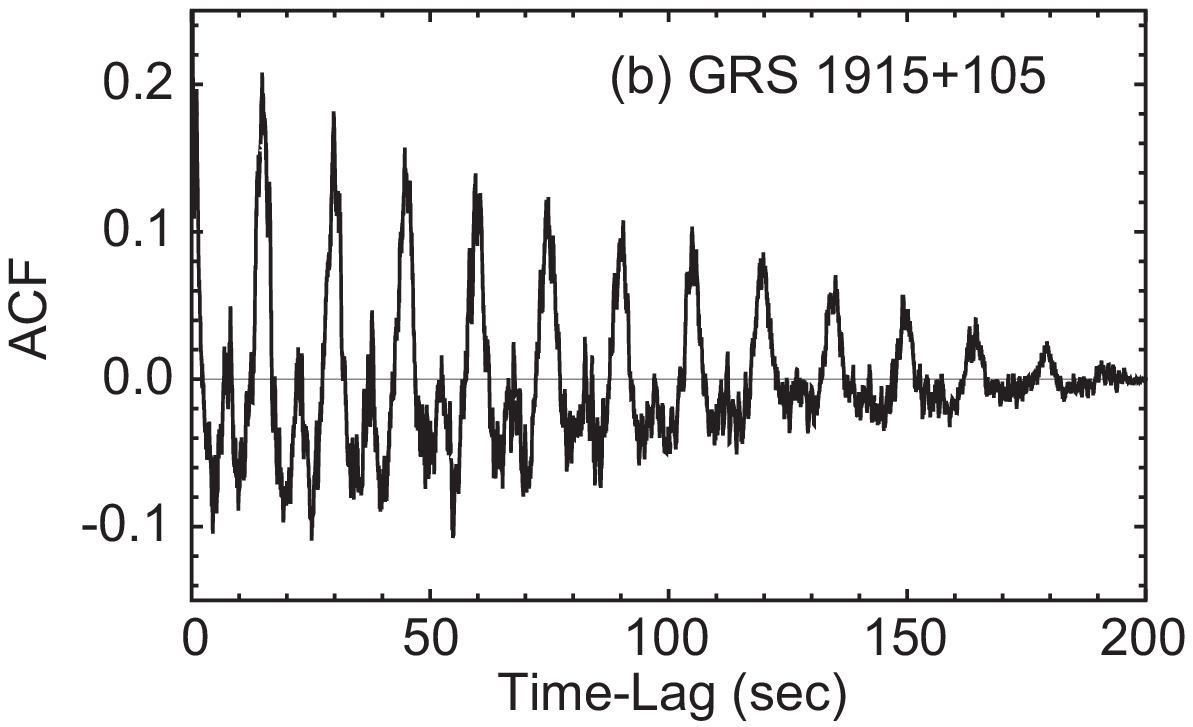}
   \caption{(a) The ACF corresponding to the model LC of
Figure~\ref{fig:f3}a showing three characteristic peaks at $\tau=0$
(self-correlation), $\backsimeq 14$ (shown with arrow) and
$\backsimeq 28$ sec. (b) The ACF corresponding to the light curve
of Figure~\ref{fig:f3}b for GRS~1915+105 (similar to
Fig.~\ref{fig:acf}c except for only $T=200$ sec here). While the
addition of noise can make both LCs look random, their ACFs are
fundamentally different.}
              \label{fig:f4}%
    \end{figure*}

The normalized PDS derived from these LCs are shown respectively in
Figure~\ref{fig:f3}c and d. Their broad band form is different
considering the Poisson character of the first LC and the $\propto
1/\nu$ of the second; however, despite the fundamentally different
apperance, they both exhibit harmonically spaced QPOs at
approximately the same frequencies (albeit of different Q-values, a
fact not surprising considering the completely incoherent character
of the first LC). According to conventional wisdom, based on the PDS
shown in Figure~\ref{fig:f3}c one would suspect a hidden periodicity
within the LC of Figure~\ref{fig:f3}a. This is indeed the case with
many QPOs of GBHCs or LMXBs, but from the algorithm used to produce
it we know there is none!.

With this in mind we present the corresponding ACFs of both LCs in
Figure~\ref{fig:f4}. It can be seen these are drastically different;
in fact the ACF of the first light curve is different from of the
ACFs of any of the sources presented in this paper despite the
presence of strong QPO in all: it consists of two narrow peaks at
$\tau = 0$ (self-correlation) and $\tau \simeq 14$ sec, a well known
characteristic of an echo in the signal (Fukumura \& Kazanas~2008,
Fukumura, Kazanas, \& Stephenson~2009, albeit of a lag much larger
than the one specific to these references). The ACF of GRS~1915+105
does exhibit a (clearly non-sinusoidal) oscillation at the
fundamental QPO period (as one would expect from an oscillating
source) but, as discussed above, it decreases linearly with time to
zero at a lag equal to the LC duration; this {\it linear} decrease
is a property of an oscillation with a phase that is preserved over
the observation interval (e.g. Morgan, Remillard, \& Greiner~1997), as one can easily attest by computing the
ACF of a simple sinusoid.

\begin{figure*}
   \centering
   \includegraphics[trim=0in 0in 0in
0in,keepaspectratio=false,width=2.7in,angle=0]{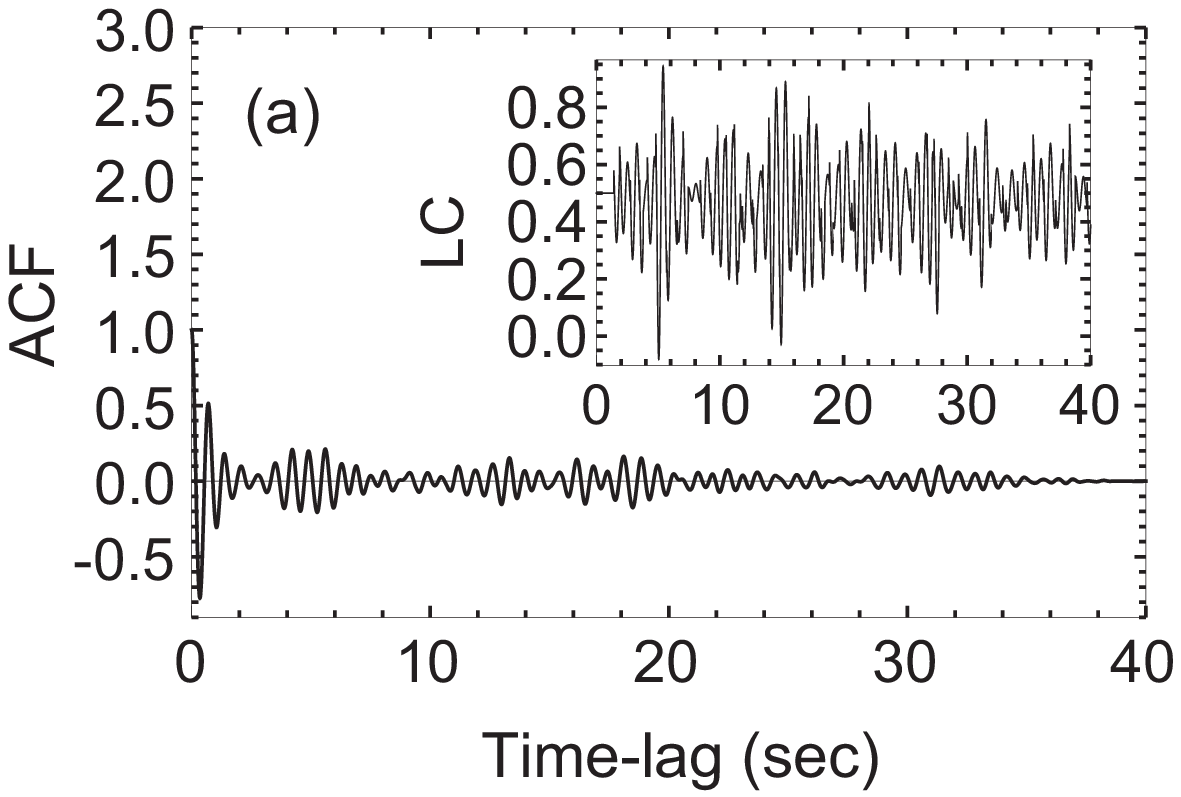}
   \includegraphics[trim=0in 0in 0in
0in,keepaspectratio=false,width=2.7in,angle=0]{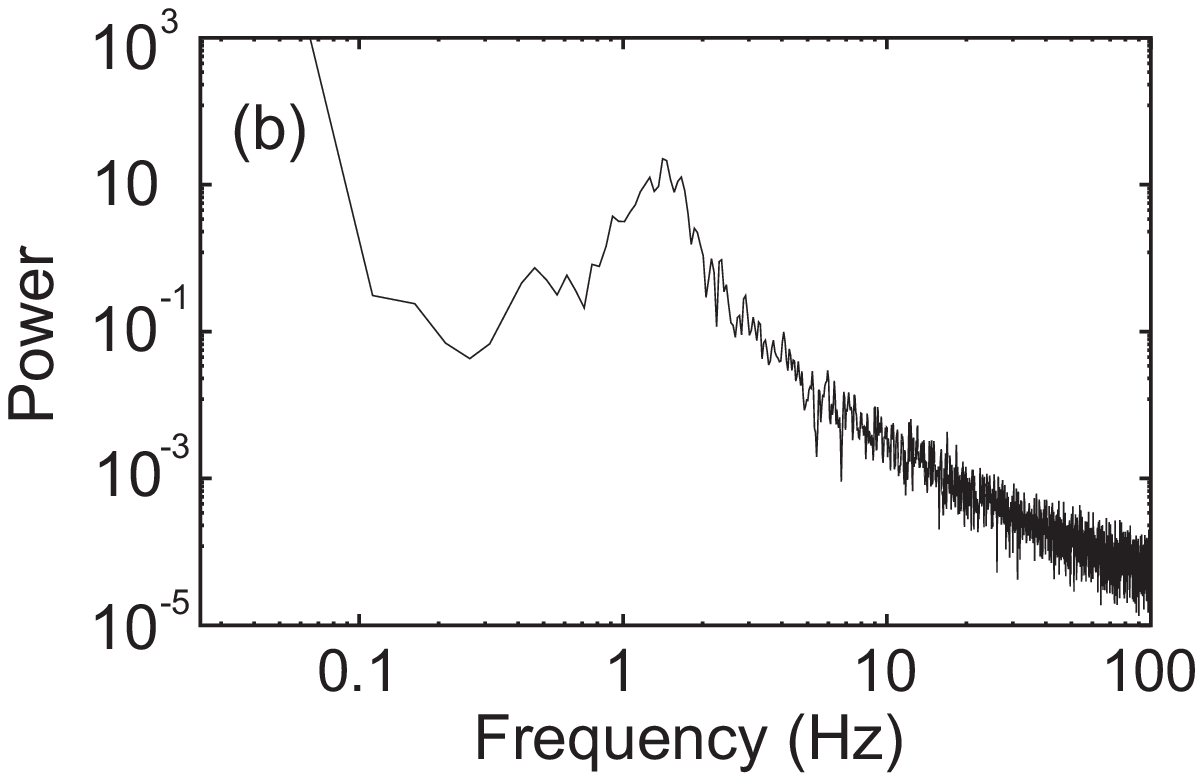} \\
   \includegraphics[trim=0in 0in 0in
0in,keepaspectratio=false,width=2.7in,angle=0]{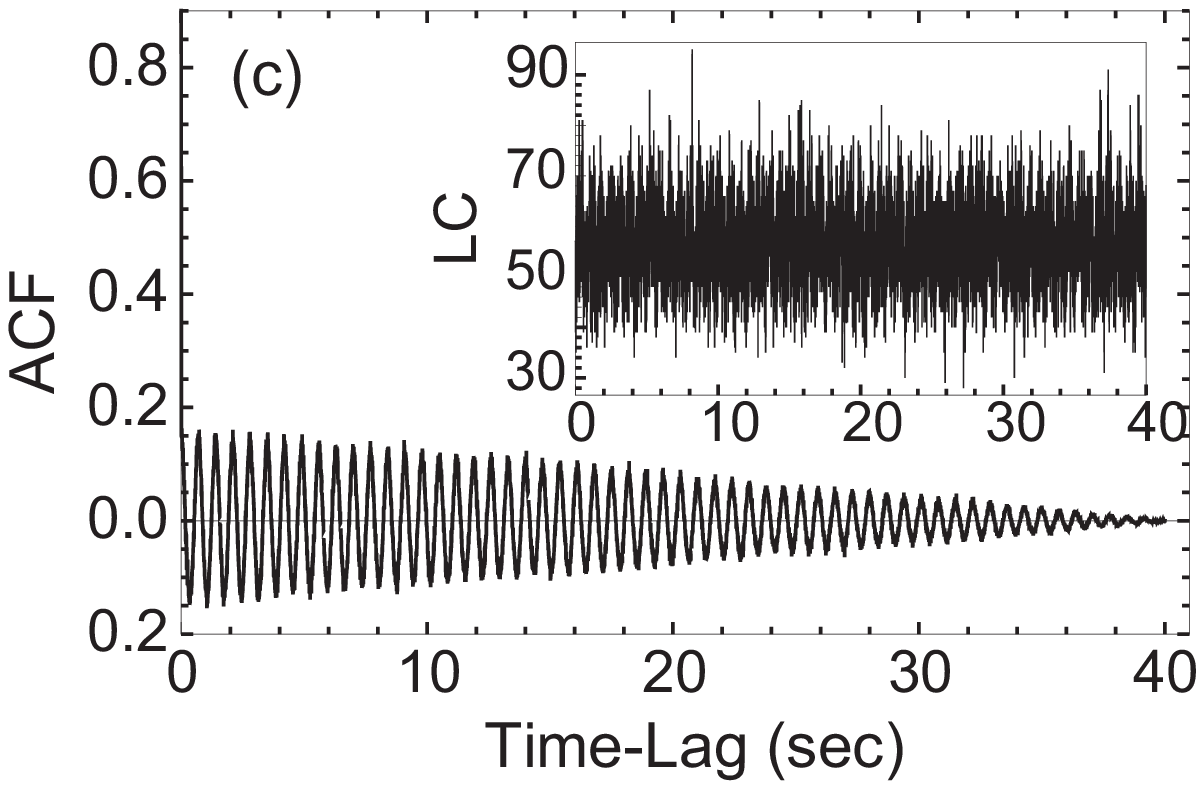}
   \includegraphics[trim=0in 0in 0in
0in,keepaspectratio=false,width=2.7in,angle=0]{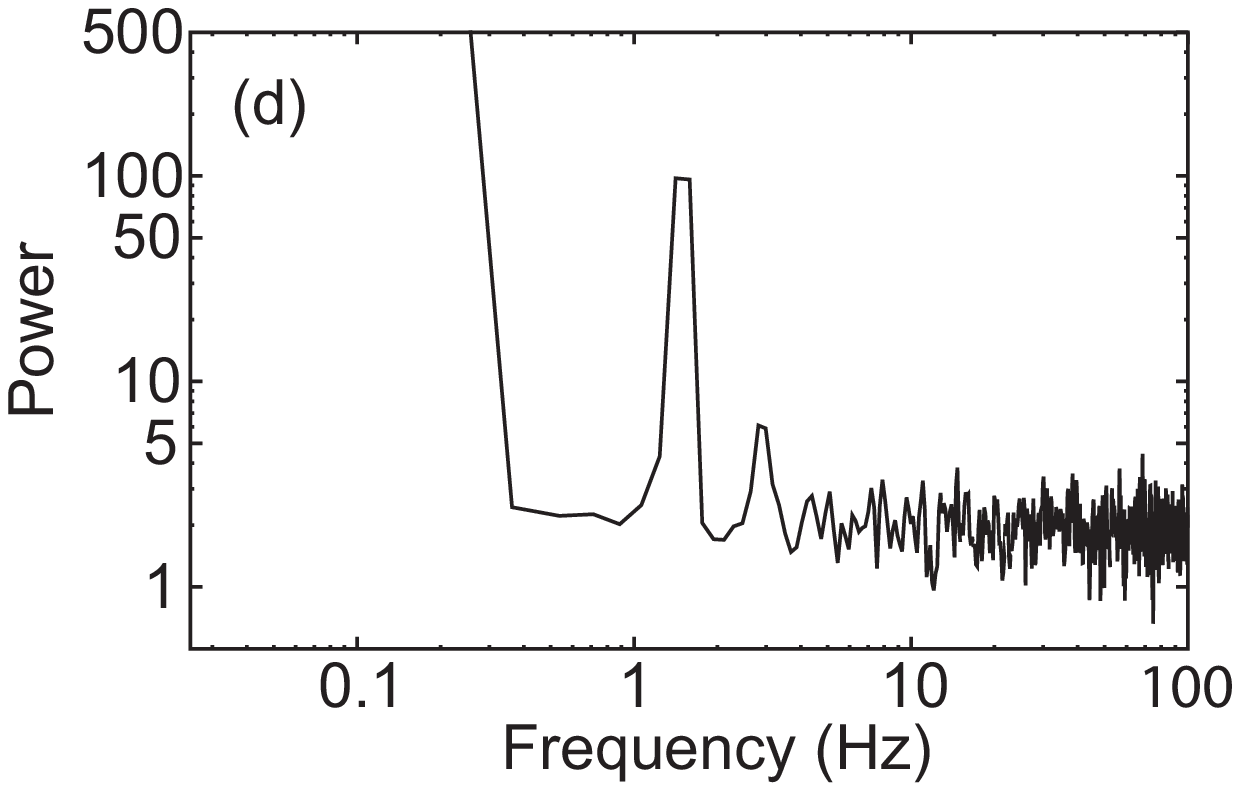}
   \caption{Modeled ACFs (left panels) and their 40sec-long LCs (insets)
   for a random-phase ensemble of $\sim 100$ damped-sinusoidal waves
   (upper panels) and $|\sin (\omega t)|$ function (lower panels)
   with period of $0.7$ sec. The corresponding PDS (right panels)
   showing prominent QPO peaks at $\sim 1.4$ Hz as expected. The damping time
   in (a)-(b) is $\sim 10$ sec. (b):
   }
              \label{fig:sine}%
    \end{figure*}

Motivated by the preservation of the oscillatory phase as the main
ingredient behind the differences in the ACFs between GRS 1915+105
and the rest of the sources
%
%
we present below two further examples of model LCs
along with their ACF and PDS that provide support to this view and
presumably shed some light into the nature of the oscillations
associated with the observed QPOs in accretion powered sources. In
Figure \ref{fig:sine}a we present a LC (40-sec long) by randomly
adding $\sim 100$ damped oscillations of the form $\propto
\sin(\omega t) \cdot e^{(-t/t_0)^2}$ [where $\omega \simeq 0.75$ Hz
is the angular frequency and $t_0 \sim 5 (2\pi/ \omega)$ is the
e-folding decay time scale] with random phases (inset) along with
its ACF (main figure); the ACF has a form very similar to those
of three of the sources in our sample, including the long time, low
amplitude oscillations. In Figure \ref{fig:sine}c present an
oscillation of the form $\propto \vert \sin(\omega t) \vert$ over 40
sec with a mean rate of 60 cts/sec along with the relevant Poisson
noise (inset); the corresponding ACF is shown in the main figure
which (modulo the single rather than double peaked oscillation)
bears  great resemblance to that of GRS~1915+105: its amplitude
drops to 0.2 after one period (the result of the presence of Poisson
noise) and subsequently decreases linearly to zero at the length of
the LC, as is the case with the ACF of GRS~1915+105. Figures
\ref{fig:sine}d present the corresponding PDS in which QPOs are
apparent, with high Q-values and higher harmonics for the case of
the LC that preserves the coherence of the oscillation phase.

As an additional argument for the complementary nature of the time
and frequency domain analysis, we note that the long term
oscillatory nature of the ACF in three of our sources, i.e. those of
the GBHCs, while obvious in the ACF, is in no way apparent in the
corresponding low frequency part of the PDS. One could argue that
the high Q-values of the QPOs of GRS~1915+105 may be indicative of
the phase coherence of the underlying oscillation; while this
property is captured quantitatively by the ACF it is impervious to
the PDS which erases all phase information. In the same vein, the
difference in the long term ACF behavior between Cyg X-2 (totally
aperiodic) and those of XTE~J1550-564 and XTE~J1859+226 (essentially
periodic oscillations) is also absent if one restricts oneself to
study of the PDS alone.



\section{Discussion \& Conclusions}

We have presented above a systematic, combined time-frequency study
of the variability properties of four galactic sources known to
exhibit QPO features. While our sample is rather limited, it
indicates that the ACFs could prove useful in providing insights
into the variability of these objects in addition to those of their
PDS. Our analysis makes clear that the properties of the
oscillations leading to the 67 mHz QPO of GRS~1915+105 are very
different from those producing the dominant QPO peaks of the other
sources, in that they continue with essentially the same phase for
the duration of the observation (the random walk in phase by $\sim
5$ sec over the 1000 sec LC segment discussed by Morgan, Remillard, \& Greiner~1997 is too small to interfere with the overall ACF form and we
believe it to be the source of the observed QPO width). By contrast
the ACF of the other three sources do resemble those of a damped
oscillator, indicating that, whatever the underlying oscillation, it
occurs in trains of limited duration $\Delta \tau \simeq 1/\Delta
\nu$ with phase random beyond this time interval and perhaps for
this reason not discernible in the LC. 

Viewed in greater detail, there are several points to note in the
ACF morphology: The ACFs exhibit their first major peak at lag $\tau
= 1/\nu_f, ~\nu_f$ being the QPO fundamental frequency. This could
serve to distinguish between the fundamental and its subharmonic in
XTE~J1550-564, which occurs close to the second ACF maximum. The
frequency of the first QPO harmonic, easily discernible in the PDS
of all sources, associates in the three sources other than GRS~1915+105,
with the first ACF minimum. In this very different
source, it associates with a secondary, albeit small amplitude,
maximum with no ACF features corresponding to the second or third
harmonics of its PDS. Furthermore, the ACF of GRS~1915+105 exhibits
an additional peak at $\tau \simeq 1$ sec (inset in Fig.
\ref{fig:acf}c), which presumably corresponds to the broad QPO at
$\nu \simeq 1$ Hz in the PDS. This is of interest because it is not
accompanied by additional oscillatory peaks as is the case with the
ACFs of the other sources. As such, it is unlikely that it
represents an independent oscillation of the type observed in the
ACFs of the other three sources and for this reason it may deserve
closer attention.

Considering the ACFs of the other three sources, they all appear
consistent with a damped oscillation. However, there are a number of
differences between those of the GBHCs and the LMXB accreting
neutron star Cyg X-2: In the GBHC ACFs the damped oscillation is
almost symmetric about zero, while in Cyg X-2 remains largely
positive. Furthermore, at lags $\tau \gg 1/\Delta \nu$ the black
hole ACFs continue to exhibit a small amplitude but {\em persistent}
oscillatory behavior that lasts through the lag span of the entire
observation. By contrast, that of Cyg X-2 exhibits nothing but
statistical fluctuations with no indication of  an oscillatory
behavior, perhaps because of input from a non oscillating source
such as the neutron star boundary layer.


Our analysis and simple model light curve examples has shown the ACF
to provide important additional information beyond that of the PDS.
This may in certain cases prove useful in figuring out the
underlying source of periodicity and its long term properties,
something not apparent in the PDS. We also provided an example in
which a strictly random source would produce QPO features in the
power spectra, whose nature while apparent in the ACF is completely
hidden in the PDS. Most of the sources we have examined have ACFs
consistent with the simplest QPO model, that of a damped
oscillations of random phases. We do not know if this is a general
property of these sources. In face of our limited sample,
we do not want to prejudge the outcome of such analyses of much
larger samples nor the efficacy of simple models and
considerations such as those discussed just above in providing
successful interpretations of the data. We anticipate, however, they
will be the subject of future publications.

\begin{acknowledgements}
The authors would like to thank the anonymous referee for a number
of useful suggestions that helped improve an earlier version of the
paper. They would also like to thank to Stratos Boutloukos, Fotis
Gavriil, Nickolai Shaposhnikov and Tod Strohmayer for useful
discussions and comments.
\end{acknowledgements}

\end{document}